\documentclass[aps,twocolumn,floatfix,titlepage,balancelastpage,raggedbottom,amsfonts,amssymb,amsmath,showkeys,superscriptaddress]{revtex4}  
\usepackage{epsfig}
\usepackage{bm}
\usepackage{times}
\usepackage{mathptmx}
\usepackage{amsmath}
\usepackage{graphicx}
\usepackage{graphics}
\usepackage{amssymb}
\usepackage{bm}
\usepackage{multirow}
\usepackage{rotating}
\usepackage{color}
\usepackage[english]{babel}
\begin{document}
\title{Simulation of capillary infiltration into packing structures by the Lattice-Boltzmann method for the optimization of ceramic materials}

\author{Danilo Sergi}
\affiliation{University of Applied Sciences SUPSI,
The iCIMSI Research Institute,
Galleria 2, CH-6928 Manno, Switzerland}
\author{Loris Grossi}
\affiliation{University of Applied Sciences SUPSI,
The iCIMSI Research Institute,
Galleria 2, CH-6928 Manno, Switzerland}
\author{Tiziano Leidi}
\affiliation{University of Applied Sciences SUPSI,
The iCIMSI Research Institute,
Galleria 2, CH-6928 Manno, Switzerland}
\author{Alberto Ortona}
\affiliation{University of Applied Sciences SUPSI,
The iCIMSI Research Institute,
Galleria 2, CH-6928 Manno, Switzerland}

\date{\today}


\begin{abstract}
In this work we want to simulate with the Lattice-Boltzmann method in 2D the capillary infiltration into porous structures obtained
from the packing of particles. The experimental problem motivating our work is the densification of carbon preforms by reactive
melt infiltration. The aim is to determine optimization principles for the manufacturing of high-performance ceramics. Simulations
are performed for packings with varying structural properties.
Our analysis suggests that the observed slow infiltrations can be ascribed to interface dynamics.
Pinning represents the primary factor retarding fluid penetration. The mechanism
responsible for this phenomenon is analyzed in detail. When surface growth is allowed, it is found that the phenomenon of pinning
becomes stronger. Systems trying to reproduce typical experimental conditions are also investigated. It turns out
that the standard for accurate simulations is challenging. The primary obstacle to overcome for enhanced accuracy seems to be the
over-occurrence of pinning.
\end{abstract}
\maketitle

\section*{1.~~~INTRODUCTION}

Porous media are ubiquitous in science and engineering.
Typical examples ripple across a number of disciplines such as hydrology, oil recovery, materials science,
printing and solar technology, to cite but a few (Bear, 1972; Dullien, 1992; Furler et al., 2012; Goodall
\& Mortensen, 2013; Koponen et al., 1998).
In simulations, their structures are investigated according to a variety of properties (mechanical, thermal and
fluid dynamics) mainly through finite elements methods (FEM) (Alawadhi, 2010; Bohn \& Garboczi, 2003).
Capillarity is the spontaneous infiltration of a fluid into a porous medium (Alava, Dube, \& Rost, 2004; Washburn, 1921).
The adhesive forces between the liquid and solid phases are responsible for this phenomenon. Indicatively, its characteristic
length scale is the micron (de Gennes, Brochard-Wyart, \& Qu\a{'}er\a{'}e, 2004). Our interest is in the role of the porous structure for capillary
infiltration. Simulations are carried out by using the Lattice Boltzmann (LB) method
(Benzi, Succi, \& Vergassola, 1992; Chen \&  Doolen, 1998; Succi, 2009;
Sukop \& Thorne, 2010; Wolf-Gladrow, 2005). This is a numerical scheme for approximating
the hydrodynamic behavior, especially suited for problems involving complex boundaries and interface phenomena.
Its main advantage over other approaches resides in the discretization of the velocity space and a statistical
treatment of particle motion and collisions. Applications encompass disparate critical systems
(Bao et al, 2014; Ghosh, Patil, Mishra, Das, \& Das, 2012; Haghani, Rahimian, \& Taghilou, 2013;
Joshi \& Sun, 2010; Liu, Valocchi, \& Kang, 2012;
Wang \& Pan, 2008). The accuracy is in general comparable with that of studies in computational fluid dynamics
(CFD) based on FEM (Succi, 2009).

The actual problem motivating our research is the reactive infiltration of molten silicon (Si) into carbon (C) preforms
(Bougiouri, Voytovych, Rojo-Calderon, Narciso, \& Eustathopoulos, 2006; Dezellus \& Eustathopoulos, 2010;
Dezellus, Hodaj, \& Eustathopoulos, 2003; Einset, 1996, 1998; Eustathopoulos, 2015;
Eustathopoulos, Nicholas, \& Drevet, 1999; Hillig, Mehan, Morelock, DeCarlo, \& Laskow, 1975;
Israel et al., 2010; Liu, Muolo, Valenza, \& Passerone, 2010; Messner \& Chiang, 1990;
Mortensen, Drevet, \& Eustathopoulos, 1997; Voytovych, Bougiouri, Calderon, Narciso, \& Eustathopoulos, 2008).
This is a complex phenomenon associated with a wetting transition
and surface growth. A feature of this process is that Si reacts with C to form silicon carbide (SiC). Si wets SiC but
does not wet C (Bougiouri et al., 2006). Furthermore, the reaction-formed phase causes the thickening of the surface
behind the invading front (Bougiouri et al., 2006; Einset, 1996, 1998; Israel et al., 2010; Messner \& Chiang, 1990).
The interaction with the fluid
flow results in a retardation of the infiltration up to its interruption because of pore clogging. This research gains
increasing attention also in industrial practice. Liquid Si infiltration (LSI) is a common manufacturing route to
high-performance ceramics. Prominent advanced applications include armor and brake systems (Fan et al., 2008;
Krenkel \& Berndt, 2005; Roberson \& Hazell, 2003).
Significant engineering efforts concentrate on the control of the effects of reactivity. Indeed, SiC formation provides
additional hardness to the resulting ceramic material, reducing also the amount  of residual unreacted Si detrimental
for high-temperature functioning. On the other hand, pore closure can stop infiltration before full densification
is attained. The microstructure of the C preform is of course of importance for optimal impregnation, besides for the
operating behavior of the final ceramic product in specific applications (e.g., the length of reinforced C fibers)
(Aghajanian et al., 2013; Gadow, 2000; Gadow \& Speicher, 2000; Israel et al., 2010; Paik et al., 2002;
Salamone, Karandikar, Marshall, Marchant, \& Sennett, 2008).

In previous works we have studied the effects of surface growth for systems composed of single capillaries both
uniform and structured in 2D (Sergi, Grossi, Leidi, \& Ortona, 2014, 2015).
The focus of the present investigation is on making more precise
the pore configurations of particular relevance for the industrial application of LSI. The strategy consists in first
realizing porous structures approximating real porous preforms. To this end, we generate packing systems
by using particles differing in basic attributes (size, shape and distribution). In principle, these parameters
can be controlled by the selection and processing of powders. The packing structures are assembled by means of
the random sequential addition (Sherwood, 1997; Widom, 1966). This is a simple and effective algorithm allowing to reach
reasonable volume fractions for random packings. The simulations for capillary infiltration aim at highlighting the function
of specific pore and structural characteristics. The systems of this work display enhanced randomness with pores
and channels of different size and orientation inducing complex dynamics for the invading front. The significance of
this investigation resides in the possibility to translate the findings into inputs for preform preparation in order
to improve the quality of ceramic materials.

Our simulations indicate that the interface dynamics can be assumed to be responsible for small effective radii. Pinning
of the contact line is the primary process affecting capillary infiltration. Our work allows to give evidence for the mechanism
determining pinning in random systems.  Packings of particles based on bimodal size distributions turn out to provide features
reducing the strong retardation of liquid invasion associated with pinning. Larger particles introduce walls creating channels
that are filled faster while smaller particles make the pore structure more uniform. As a result, the interface travels longer
distances before the loss of its curvature. Furthermore, the probability to encounter the surface of other particles is higher.
Our data also show that the inertia effects induced by accelerations remain of secondary importance. With surface growth, it is
found that pore closure can be effective
and pinning turns out to be enhanced. The retardation effects are stronger for fast infiltrations (higher degree of orientation disorder),
that remain nevertheless more advantageous. Also with surface growth, bimodal size distributions lead to better results.
Simulations with higher resolution for the porous structure are also performed for a direct comparison with experimental
expectations. In this case, the effect of surface growth on pore closure is mainly indirect because pinning is
stronger. Furthermore, the relative strength of capillary forces is too weak. This means that the details of
the infiltration process suffers from discrepancies with real conditions. For example, this implies that the ratio of the
infiltrated distance to the thickness of the growing surface (i.e., the width of the reaction-formed phase)
is too small in simulations. Progress seem possible especially by reducing the phenomenon of pinning.


\section*{2.~~~MODELS AND ANALYSIS}

For the formalism of the LB models the interested reader is addressed to Sergi et al.~(2014). In the sequel,
the same notation and terminology are employed.  We recall that in experiments with molten Si and some alloys
the invading front exhibits a kinetics linear with time (Israel et al., 2010; Voytovych et al., 2008). The
standard Washburn law for capillary infiltration in vacuum predicts a parabolic time dependence. The observed
experimental behavior can be reproduced with the simulations by assuming that the systems are composed by two
fluid components (Chibbaro, 2008; Chibbaro, Biferale, Diotallevi, \& Succi, 2009;
Diotallevi, Biferale, Chibbaro, Lamura, et al., 2009; 
Diotallevi, Biferale, Chibbaro, Pontrelli, et al., 2009). This means that
the effect of the reaction at the contact line is equivalent to the resistance due to the presence of a second
non-wetting component as dense and viscous as the wetting one (Sergi et al., 2014). This approach does not
explain the origin of the linear behavior for the infiltration kinetics generally ascribed to effects of the
reactivity (Israel et al., 2010; Voytovych et al., 2008).

The differential equation describing capillary infiltration for two fluids with the same density $\rho$ and
dynamic viscosity $\mu$ into a single channel in 2D reads (Chibbaro, Biferale, Diotallevi, et al.,
2009)
\begin{equation}
\frac{\mathrm{d}^{2}z(t)}{\mathrm{d}t^{2}}+\frac{3\mu}{r^{2}\rho}\frac{\mathrm{d}z(t)}{\mathrm{d}t}
=\frac{\gamma\cos\theta}{r\rho L}\ .
\label{eq:diff}
\end{equation}
The first term on the left-hand side accounts for inertial forces while the second one introduces
viscous forces. The term on the right-hand side is due to capillary forces.
$z$ designates the position of the invading front, $r$ is the radius of the channel and $L$ its length.
When inertial forces are neglected, for a shrinking radius as $r(t)=r_{0}-kt$ ($r_{0}$ is the initial radius and $k$
is the reaction-rate constant) a solution to the above equation is given by
\begin{equation}
z(t)=\frac{\gamma\cos\theta}{6\mu k L}\big[r_{0}^{2}-(r_{0}-kt)^{2}\big]+z_{0}\ .
\label{eq:reac}
\end{equation}
$z_{0}$ is the initial position. For completeness, in the absence of reaction, a solution to Eq.~\ref{eq:diff}
taking into account also inertial forces is obtained by (Chibbaro, Biferale, Diotallevi, et al., 2009)
\begin{equation}
z(t)=\frac{V_{\mathrm{cap}}r\cos\theta}{3L}t_{\mathrm{d}}\big[\exp\big(-t/t_{\mathrm{d}}\big)+t/t_{\mathrm{d}}
-1\big]+z_{0}\ ,
\label{eq:linear}
\end{equation}
where $V_{\mathrm{cap}}=\gamma/\mu$ and $t_{\mathrm{d}}=\rho r^{2}/3\mu$.
For more general porous systems, the radius of the channel is replaced with the effective radius.
Relevant studies for reactive infiltration based on dynamics can be found in Asthana, Singh, and Sobczak (2005), Gern and
Kochend\a{"}orfer (1997), Martins, Olson, and Edwards (1988), Messner and Chiang (1990),
Sangsuwan, Orejas, Gatica, Tewari, and Singh (2001),
Yang and Ilegbusi (2000).

The simulations output is in model units. For the units of the basic quantities of mass, length and
time we use the symbols mu, lu and ts, respectively. The data obtained from simulations can be expressed
in regular units of the SI system after suitable transformations
(Chibbaro, Biferale, Binder, et al., 2009; Chibbaro, Costa, et al., 2009;
Gross, Varnik, Raabe, \& Steinbach, 2010; Joshi \& Sun, 2010;
Komnik, Harting, \& Herrmann, 2004). For
comparisons with experimental data we apply the approach based on dimensionless numbers like the Reynolds or capillary
numbers (Landau \& Lifshitz, 2008). For example, these numbers compare the relative strength between the
effects of inertial, viscous and capillary forces. Two systems are assumed to be equivalent when the dimensionless
numbers for the dominating forces have the same values. For our purposes, we also need dimensionless numbers
involving the characteristic time set by the surface reaction (Sergi, Camarano, Molina, Ortona, \& Narciso, 2016):
\begin{equation*}
Q_{1}=\frac{\rho rvk}{\gamma}\ ,\quad Q_{2}=\frac{\rho rk}{\mu}\ ,\quad Q_{3}=\frac{k}{v}\ .
\end{equation*}
$v$ indicates a characteristic velocity for the system.
$Q_{1}$ compares the reaction with surface tension, $Q_{2}$ the reaction with viscosity, $Q_{3}$ the reaction
with inertia. For completeness, $Q_{4}=\rho rv^{2}/\gamma$ compares inertia with capillary forces.
In our terminology, the first force becomes more important as the dimensionless number increases.
In our notation, the Reynolds and capillary numbers are given by $Re=\rho vr/\mu$ and $Ca=v\mu/\gamma$.
In the onset of pore closure, $Q_{1}$ tends toward zero faster and thus surface tension is the
dominant force. Instead, inertia has the weakest effects since $Q_{3}$ diverges while $Q_{1}$ and $Q_{2}$
tend to $0$. By using for the time-varying effective radius the average value $r_{0}/2$, we arrive at
the same conclusions.

For the characterization of the porous systems
we use Darcy law (Sukop \& Thorne, 2010):
\begin{equation*}
v\varphi=-\frac{K}{\mu}\nabla P\ .
\end{equation*}
$v$ is the interstitial velocity, $\varphi$ the porosity, $P$ the pressure and $K$ the permeability.
Another quantity that we take into account is the tortuosity (Bear, 1972). It is defined as the ratio
of the actual length of the capillary paths to the corresponding length of the sample. For the computations
we use the formula $<u>/<u_{x}>$ (Duda, Koza, \& Matyka, 2011; Matyka \& Koza, 2012), where $u$ is the magnitude
of the fluid velocity of components $u_{x}$ and $u_{y}$. It is worth recalling that at the interface
there are spurious currents (Wagner, 2003).

\begin{table}[t]
\begin{ruledtabular}
\begin{tabular}{lccccc}
Type  & Diameter & Volume ($10^{-3}$) & Filling\\
\hline
bimodal1  & $0.2/0.16$ & $4.2/2.1$ & $0.46$\\
bimodal2  & $0.2/0.12$ & $4.2/0.9$ & $0.46$\\
bimodal3  & $0.2/0.08$ & $4.2/0.3$ & $0.46$\\
multimodal& $0.2/0.16/0.12/0.08$ & $4.2/2.1/0.9/0.3$ & $0.46$\\
granular1 & $0.2$      & $4.2$ & $0.46$\\
granular2 & $0.16$     & $2.1$ & $0.46$\\
fine1     & $0.12$     & $0.9$ & $0.46$\\
fine2     & $0.08$     & $0.3$ & $0.44$\\
\end{tabular}
\end{ruledtabular}
\caption{\label{tab:sphere}
Relevant data for the packing structures composed of spheres. Multiple values for their
average size are given in the case of bimodal and multimodal mixtures. Figure
\ref{fig:section} visualizes some examples in 2D.}
\end{table}
\begin{table}[t]
\begin{ruledtabular}
\begin{tabular}{lccccc}
Type  & Side & Volume ($10^{-3}$) & Filling\\
\hline
bimodal1  & $0.2/0.16$ & $9/3.9$ & $0.44/0.45/0.38$\\
bimodal2  & $0.2/0.13$ & $9/1.9$ & $0.45/0.46/0.40$\\
bimodal3  & $0.2/0.08$ & $9/0.6$ & $0.44/0.46/0.45$\\
multimodal& $0.2/0.16/0.13/0.08$ & $9/3.9/1.9/0.6$ & $0.47/0.44/0.45$\\
granular1 & $0.2$      & $9$   & $0.44/0.43/0.34$\\
granular2 & $0.16$     & $3.9$ & $0.43/0.44/0.33$\\
fine1     & $0.13$     & $1.9$ & $0.41/0.42/0.33$\\
fine2     & $0.08$     & $0.6$ & $0.39/0.38/0.34$\\
\end{tabular}
\end{ruledtabular}
\caption{\label{tab:rhomb}
Basic information for the porous systems obtained from the packing of rhombs. The height of
the rhombs is comparable to the length of their four equal sides, i.e.~$h=0.2,0.16,0.12,0.08$. Bimodal
and multimodal mixtures are characterized by multiple data for the size of the particles. The values
for the filling fraction correspond to different orientations of the particles: aligned, misaligned,
misaligned and tilted, respectively. Examples for the porous structures in 2D are shown in
Fig.~\ref{fig:section}.}
\end{table}
\begin{table}[b]
\begin{ruledtabular}
\begin{tabular}{lccccc}
Type   & Side & Volume ($10^{-3}$) & Filling\\
\hline
fiber1 & $0.28/0.2$  & $1/9$     & $0.39/0.40/0.41$\\
fiber2 & $0.28/0.16$ & $1/3.9$   & $0.40/0.41/0.39$\\
fiber3 & $0.28/0.13$ & $1/1.9$   & $0.41/0.40/0.37$\\
fiber4 & $0.24/0.2$  & $0.9/9$   & $0.39/0.40/0.41$\\
fiber5 & $0.24/0.16$ & $0.9/3.9$ & $0.40/0.41/0.39$\\
fiber6 & $0.24/0.13$ & $0.9/1.9$ & $0.41/0.42/0.37$\\
fiber7 & $0.2/0.2$   & $0.7/9$   & $0.40/0.40/0.41$\\
fiber8 & $0.2/0.16$  & $0.7/3.9$ & $0.40/0.41/0.40$\\
fiber9 & $0.2/0.13$  & $0.7/1.9$ & $0.41/0.41/0.37$\\
\end{tabular}
\end{ruledtabular}
\caption{\label{tab:fiber}
Data for the packing structures obtained with fibers and rhombs. For the side and the
volume, the first value refers to the sticks and the second to the rhombs. For fibers,
the side of the square base is $0.06$ long. The height of the rhombs is $h=0.2,0.16,0.12,0.08$,
that is, similar to the length of the four equal sides. The reported filling fractions are
associated with different orientations of the particles: aligned, misaligned, misaligned and
tilted, respectively. Figure \ref{fig:section} shows examples of porous configurations in 2D.}
\end{table}
\begin{figure}[t]
\includegraphics[width=4cm]{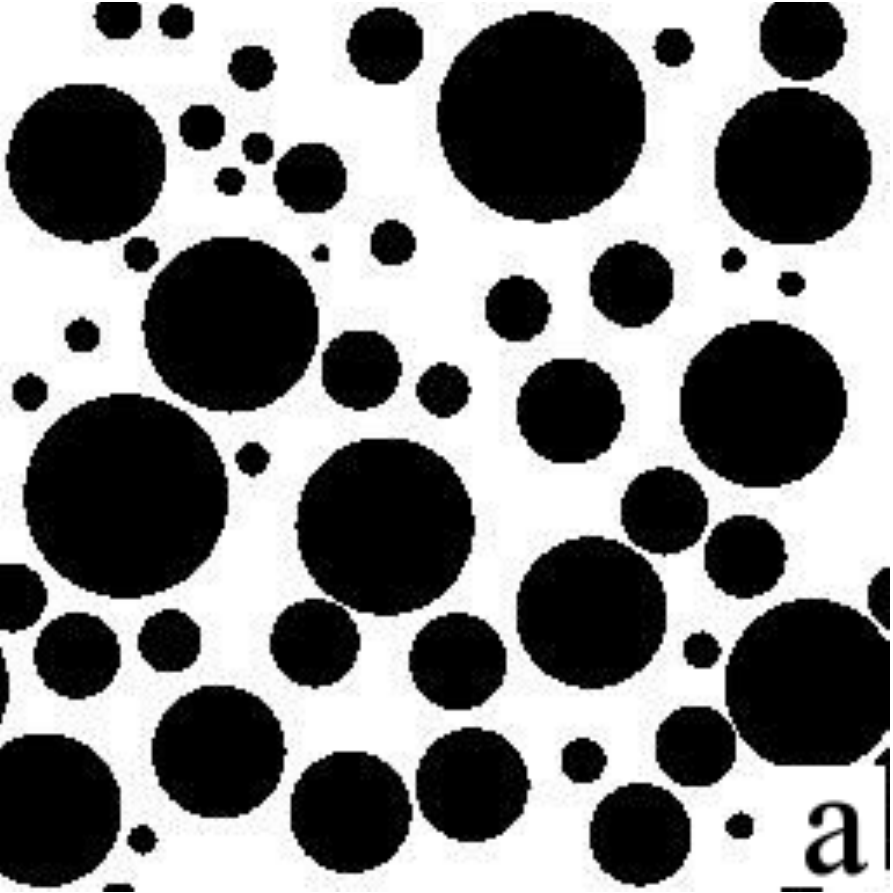}\hspace{0.4cm}
\includegraphics[width=4cm]{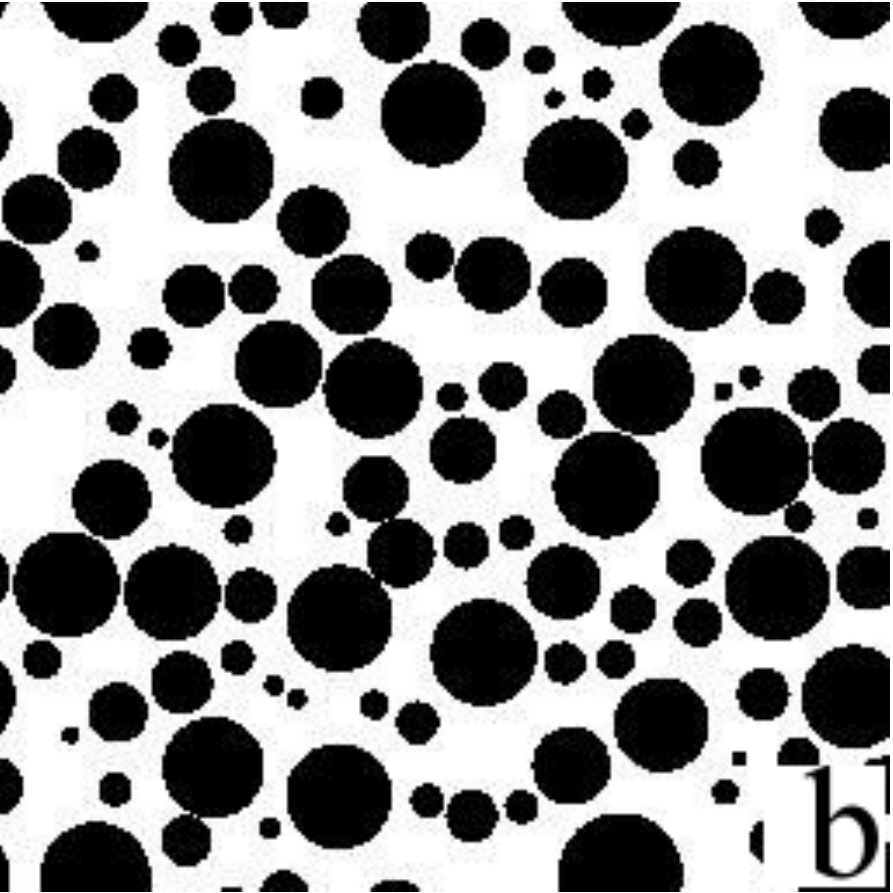}\\
\includegraphics[width=4cm]{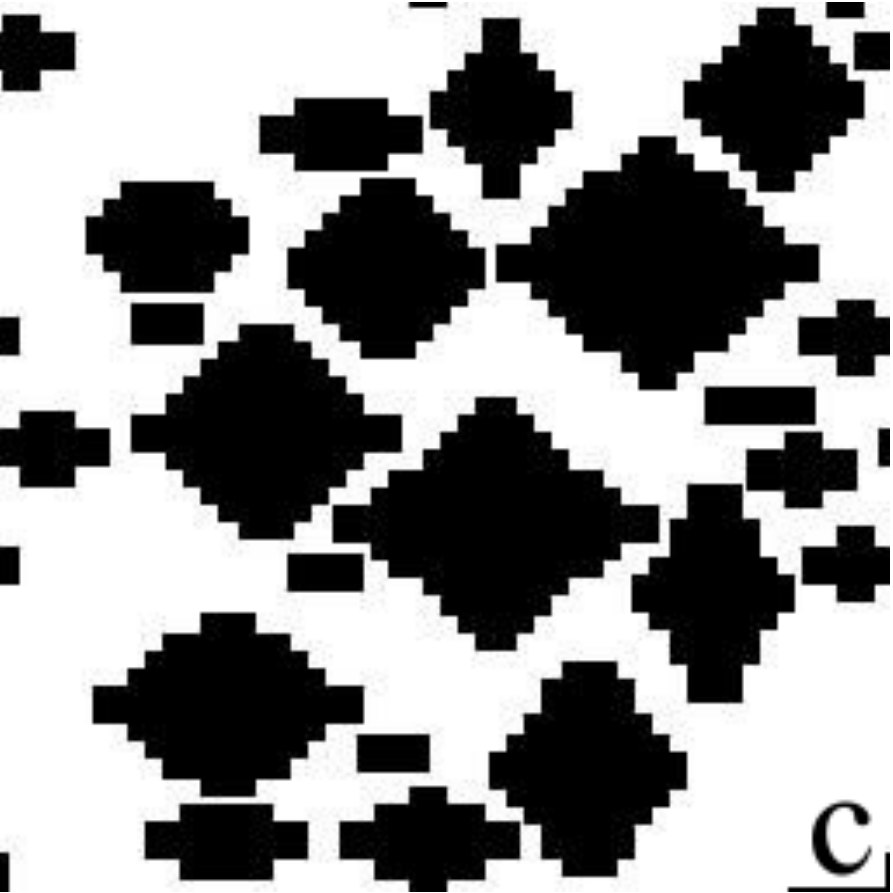}\hspace{0.4cm}
\includegraphics[width=4cm]{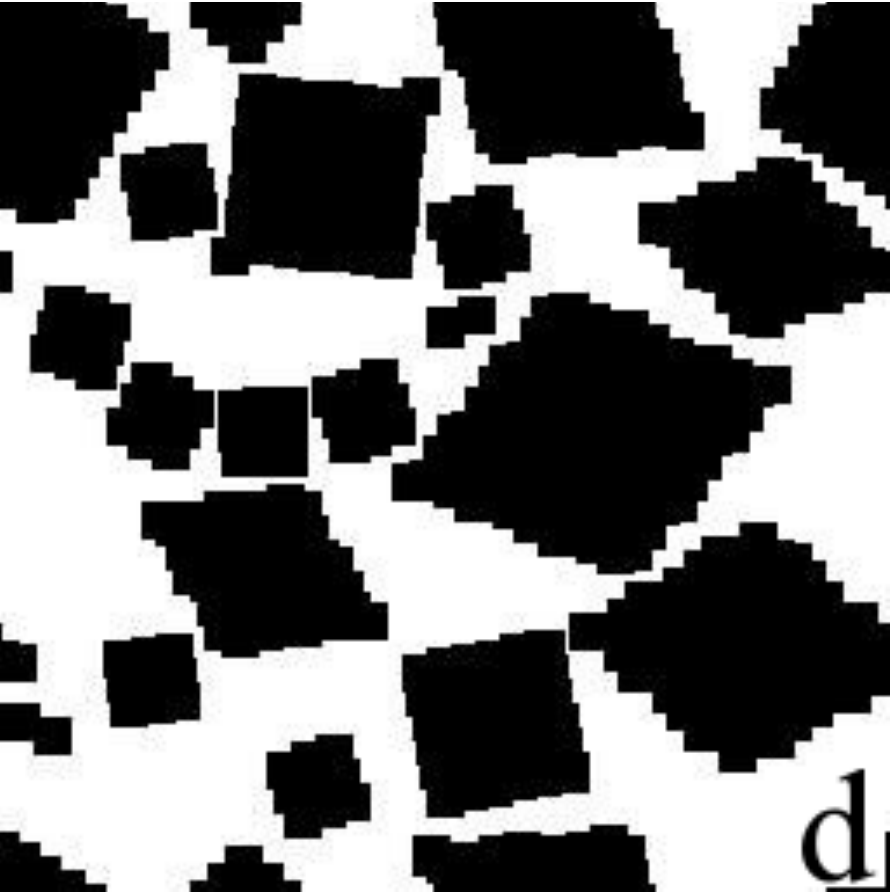}\\
\includegraphics[width=4cm]{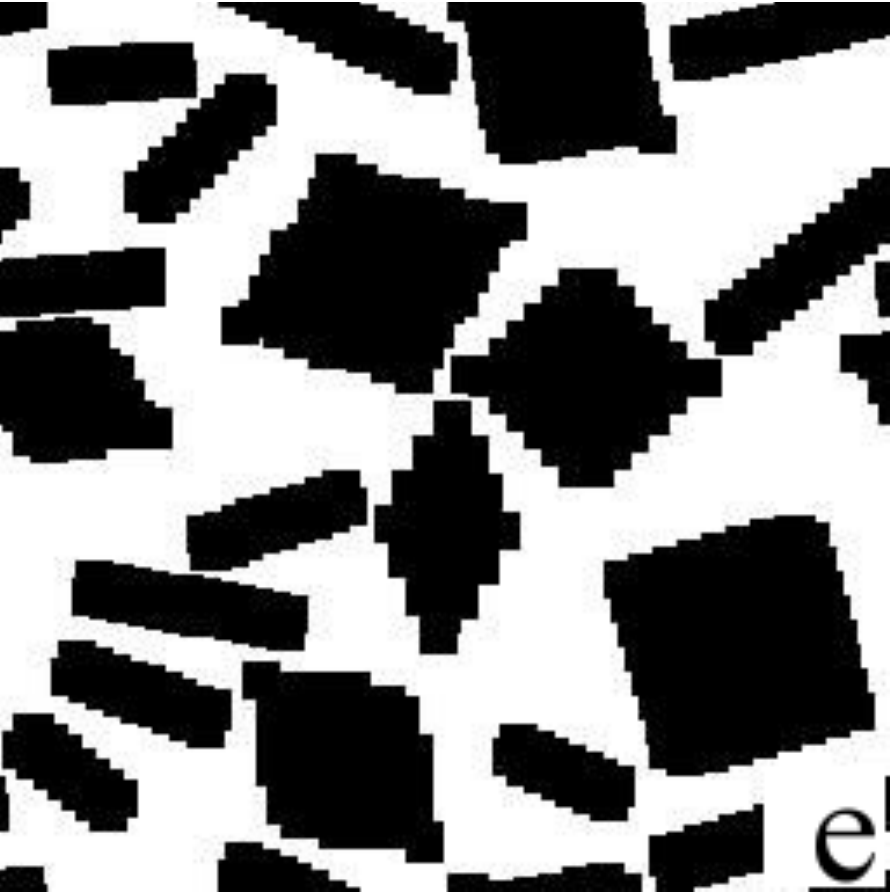}\hspace{0.4cm}
\includegraphics[width=4cm]{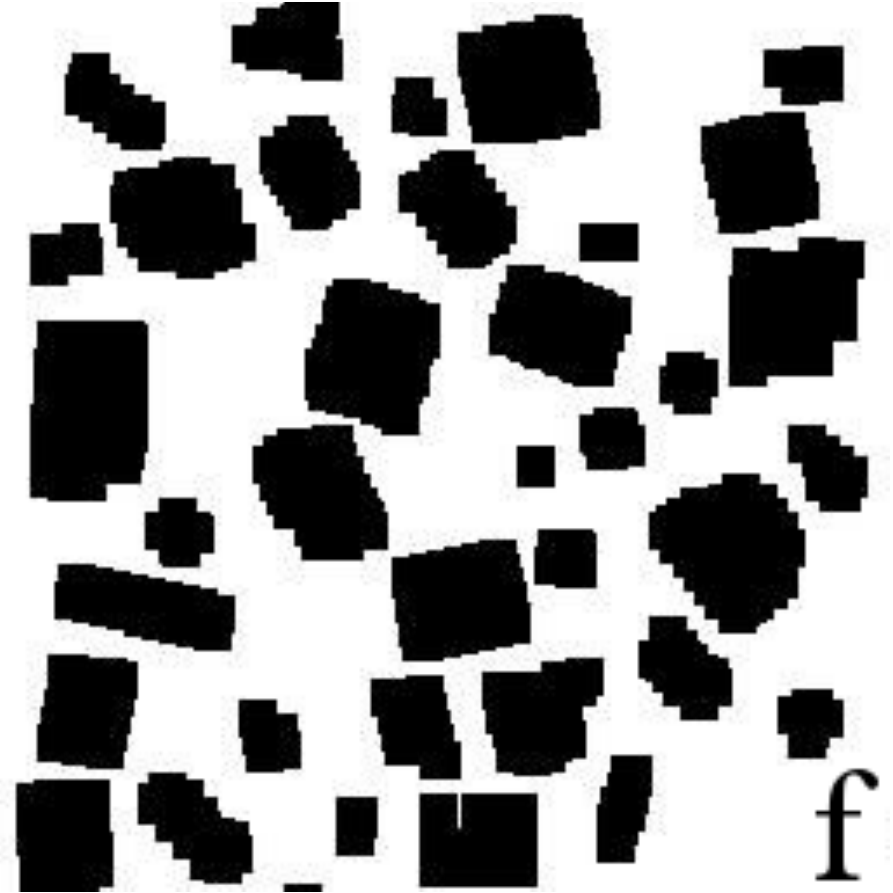}\\
\includegraphics[width=8.4cm]{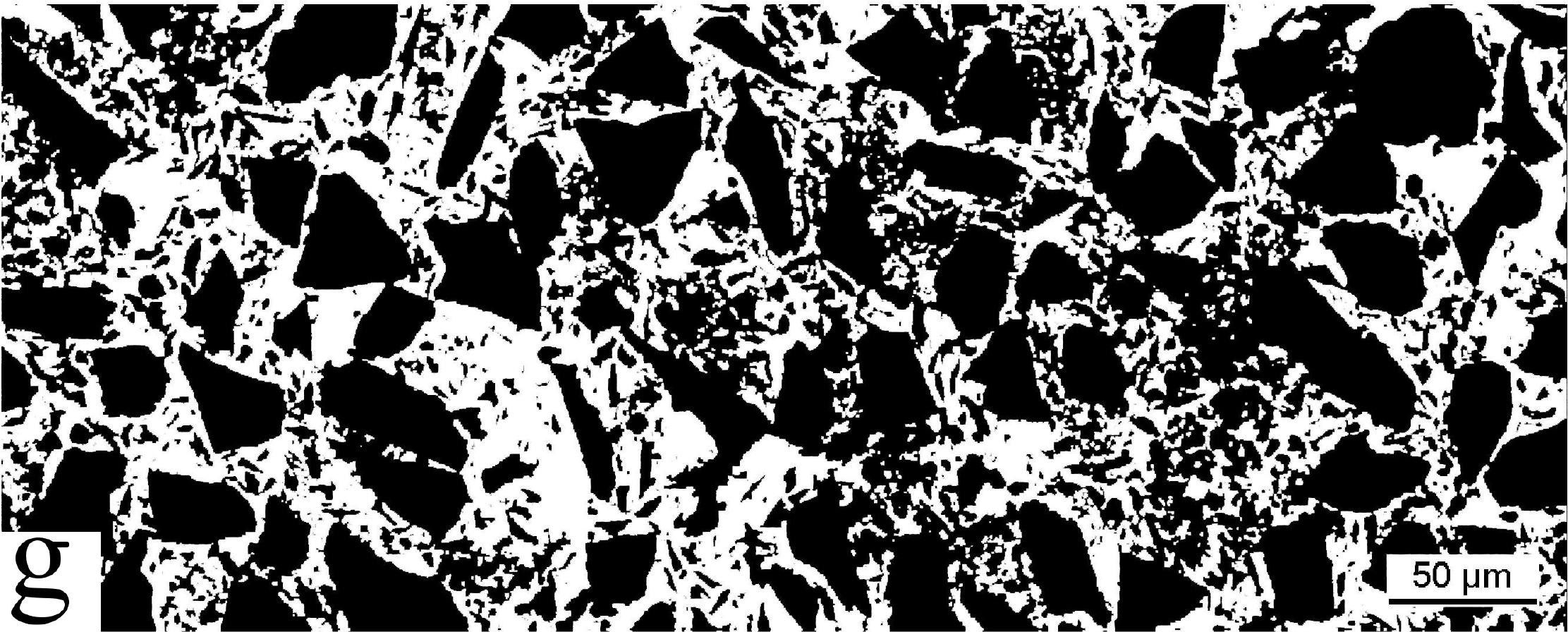}
\caption{\label{fig:section}
2D sections for packing structures; in dark the filler particles. Information on the whole structures
can be found in Tabs.~\ref{tab:sphere}-\ref{tab:fiber}. (a) System bimodal1 for spheres;
(b) system fine1 for spheres; (c) system multimodal for blocks; (d) system bimodal3 with
misalignment for blocks; (e) system fiber1 for misalignment; (f) system fiber6
for misalignment and tilt. (g) Microstructure of a typical preform.}
\end{figure}
\begin{figure}[t]
\includegraphics[width=8.4cm]{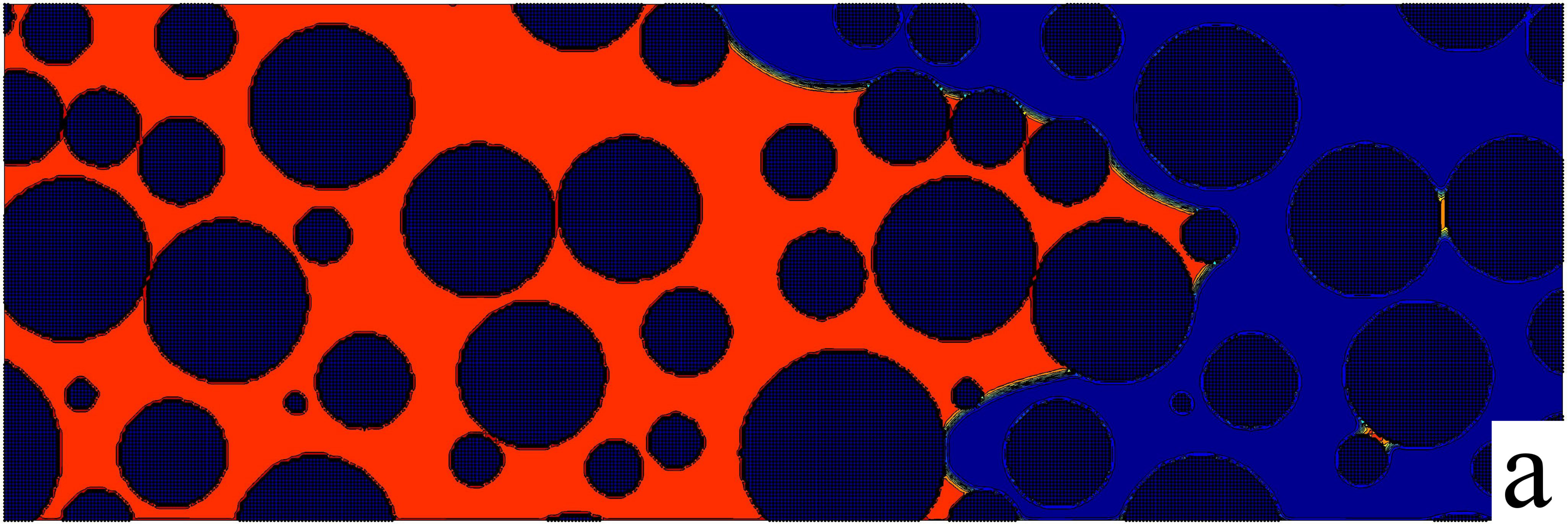}\\
\includegraphics[width=8.4cm]{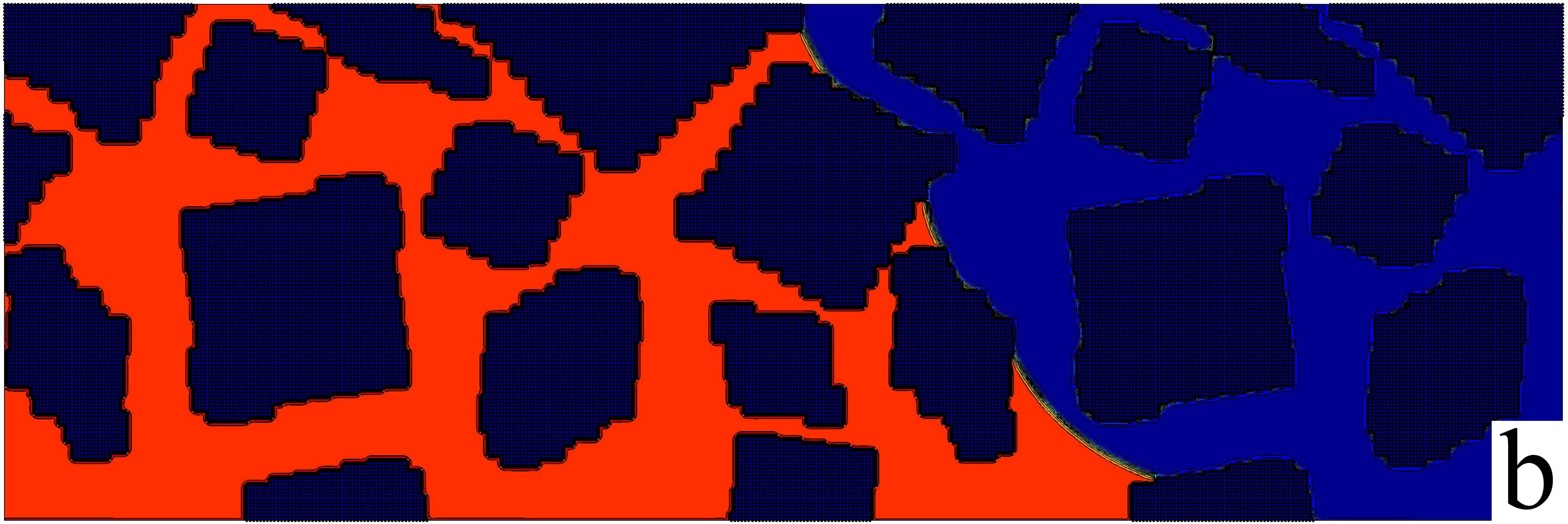}\\
\includegraphics[width=8.4cm]{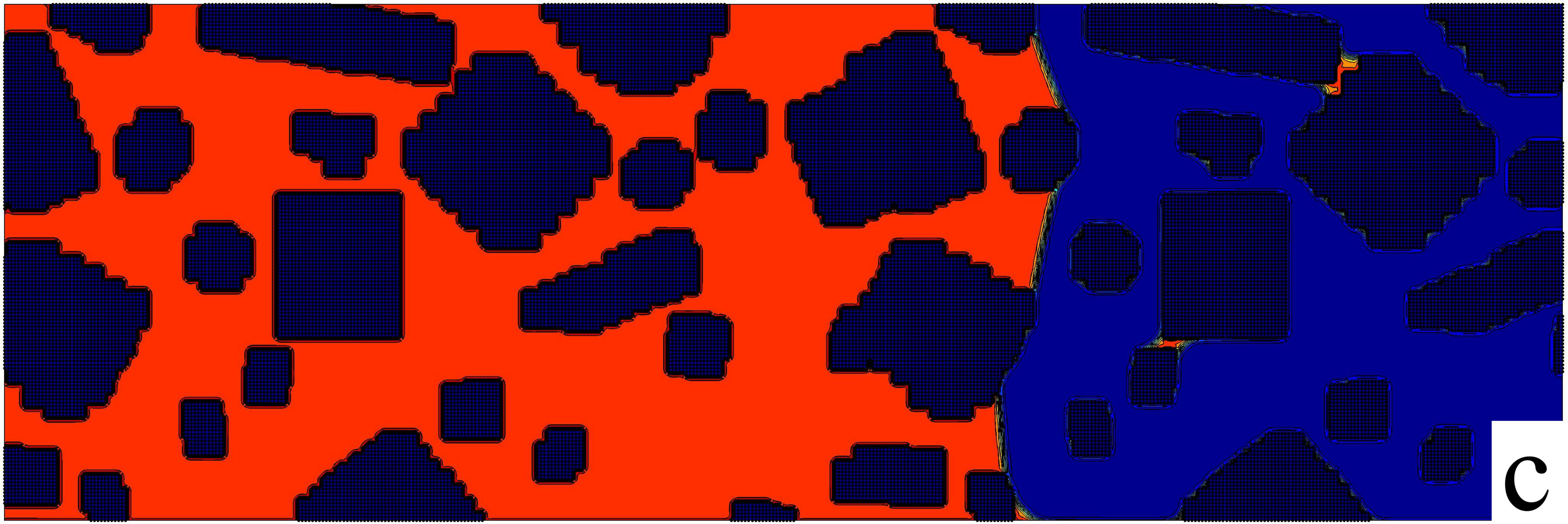}\\
\includegraphics[width=8.4cm]{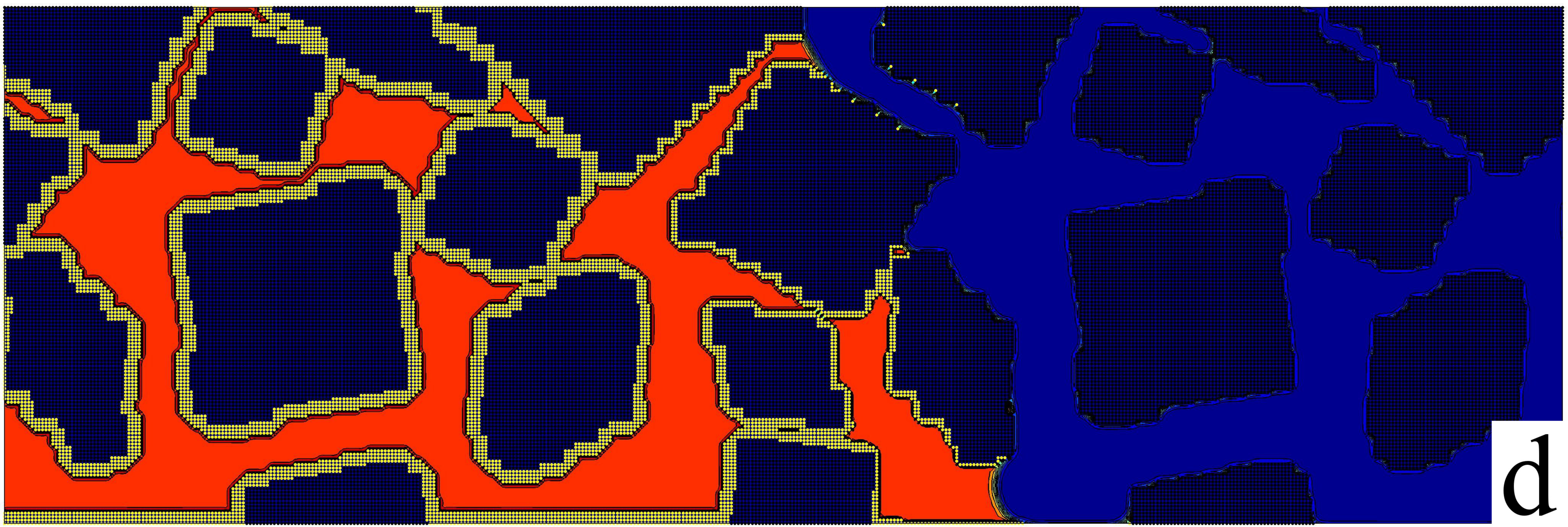}
\caption{\label{fig:front}
Fluid invasion for selected packing structures after $10^{6}$ ts. Red is used for the wetting fluid and blue for the non-wetting component.
The initial solid phase is represented in dark and the growing surface from the reaction in yellow. More details on the properties of the packing
systems can be found in Tabs.~\ref{tab:sphere}-\ref{tab:fiber}. More details on the results of infiltrations are reported in
Tabs.~\ref{tab:sphere_result}-\ref{tab:reaction_result}. (a) Simulation for the system bimodal1 with spherical particles. The formation of a
finger can be recognized. (b) Infiltration for the system bimodal2 with rhombs in the presence of misalignment. The front will further advance because
it can hit the corner of another particle before loosing its curvature. (c) Visualization of the system with fibers of type fiber2 with misalignment and
tilt. Tilted fibers result in the presence of small grains in 2D sections favoring infiltration as for bimodal size distributions. (d) Simulation for the
system bimodal2 for rhombs with misalignment and surface reaction enabled (cf.~panel b). Infiltration stops because of pore closure.}
\end{figure}
\begin{figure}[t]
\includegraphics[width=8.4cm]{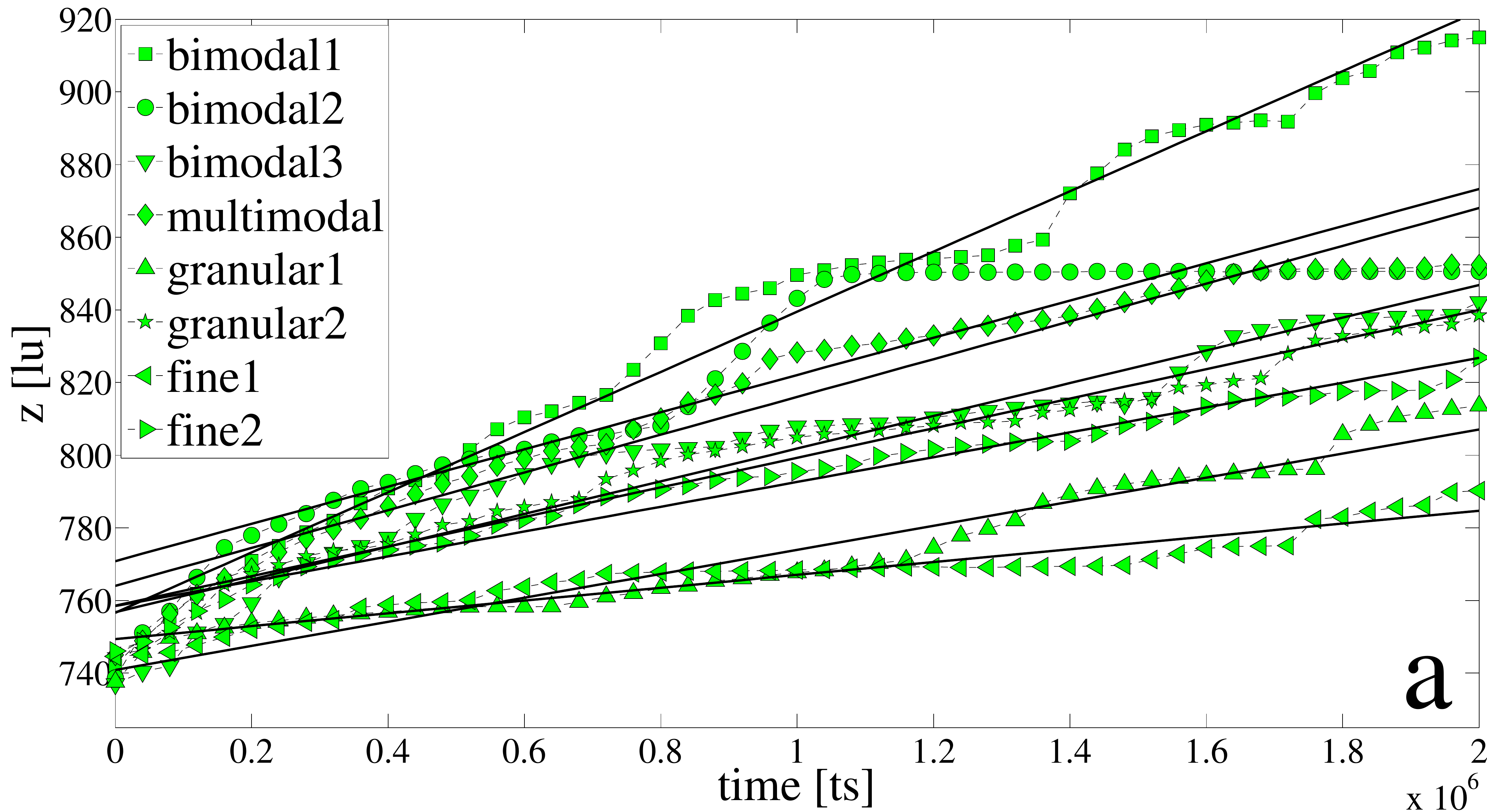}\\
\includegraphics[width=8.4cm]{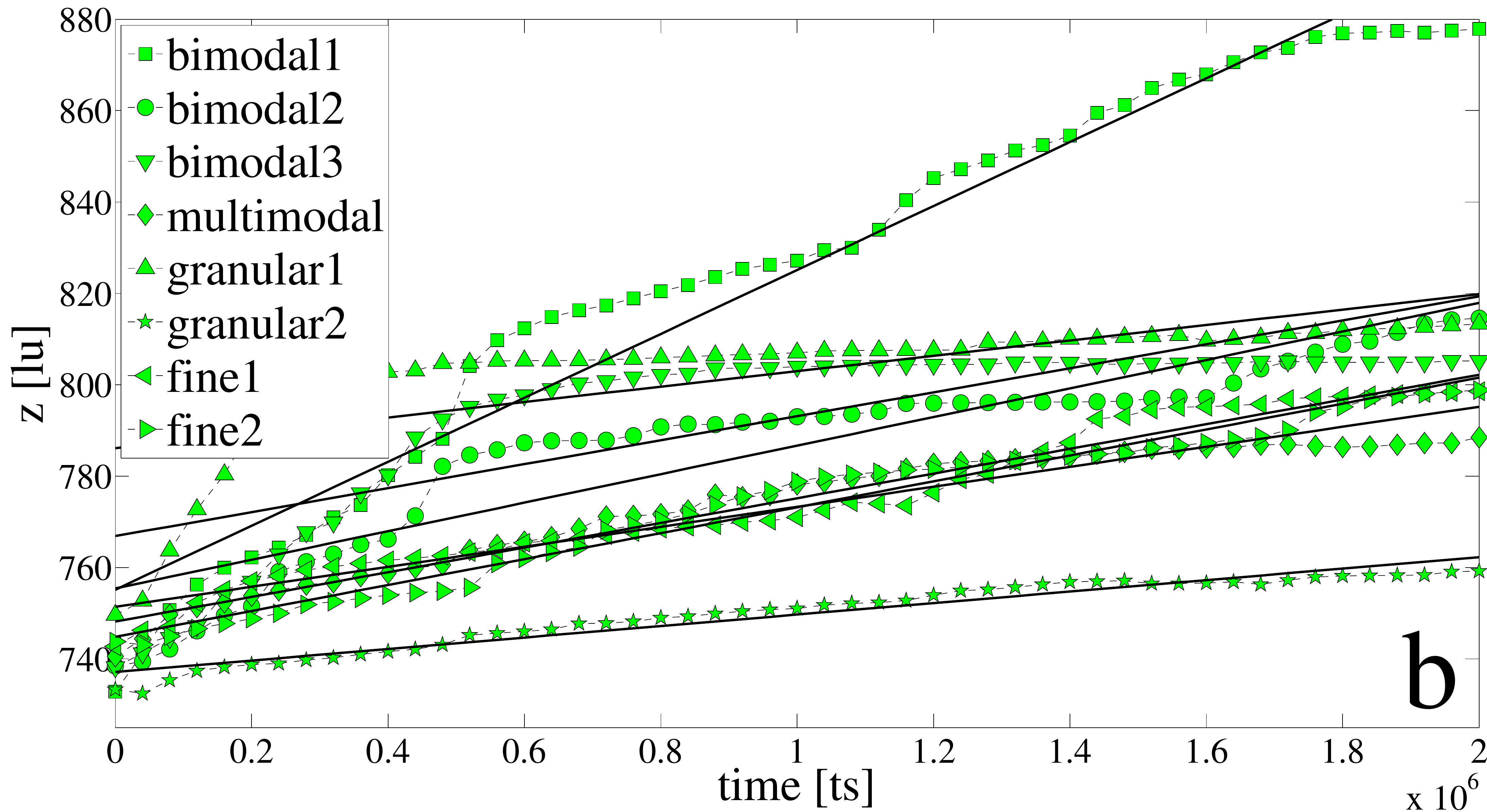}
\caption{\label{fig:filling1}
Infiltrated distance with chemical inertness for the solid surface.
Points represent simulation results. The solid lines are fits to the
data using Eq.~\ref{eq:linear}. (a) Packing structures obtained from spheres;
(b) packing structures realized with aligned rhombs. Basic properties of the
packing systems are listed in Tabs.~\ref{tab:sphere} and \ref{tab:rhomb}.}
\end{figure}
\begin{figure}[t]
\includegraphics[width=8.4cm]{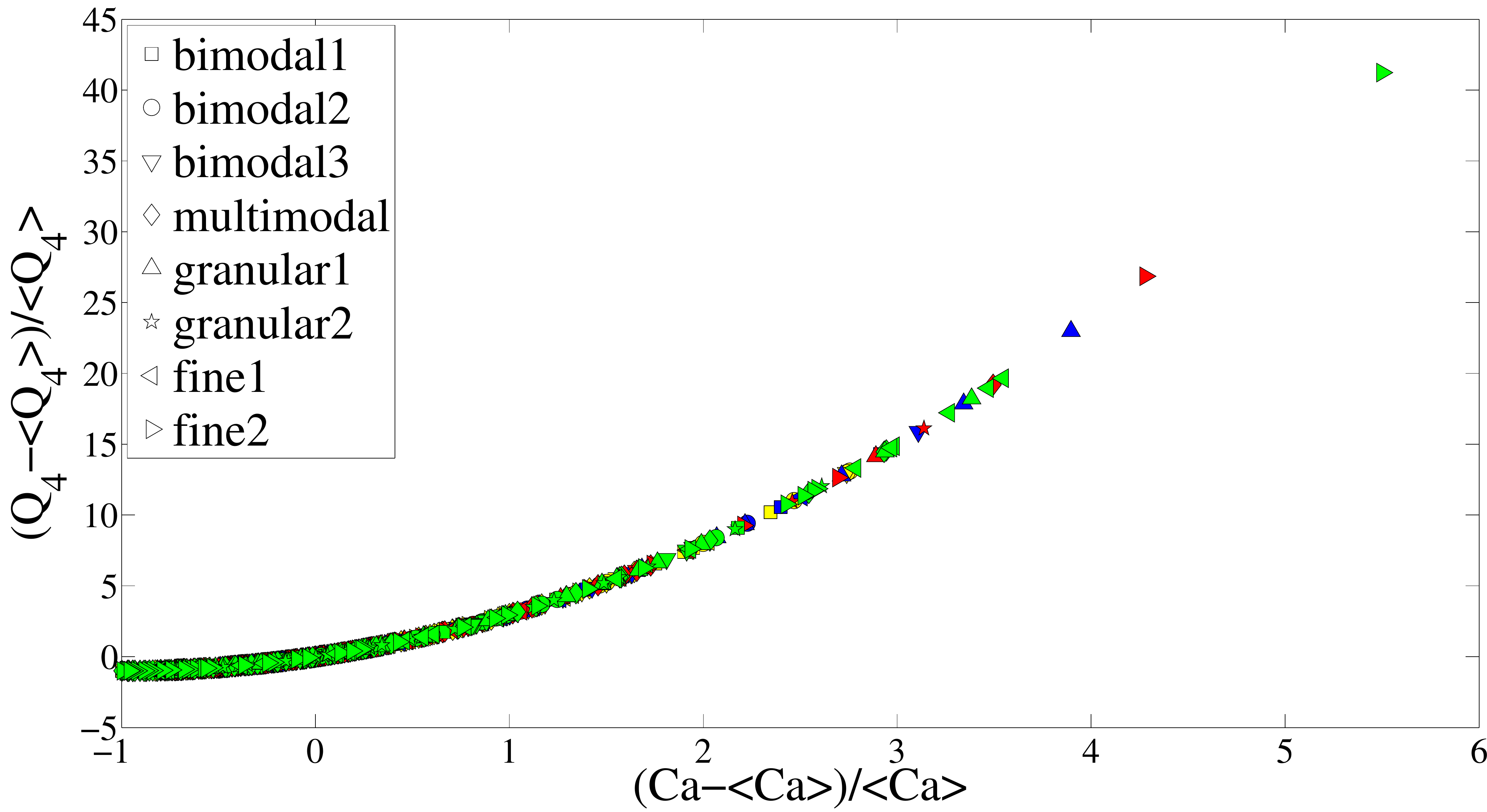}
\caption{\label{fig:q4}
Relationship between the fluctuations of characteristic numbers $Q_{4}$ and $Ca$ for packing systems with
spheres and rhombs in the absence of surface reaction. Yellow markers are used for spheres, blue for aligned
rhombs, red for rhombs with misalignment and green for rhombs with misalignment and tilt
(see Tabs.~\ref{tab:sphere} and \ref{tab:rhomb}).}
\end{figure}
\begin{figure}[t]
\includegraphics[width=8.4cm]{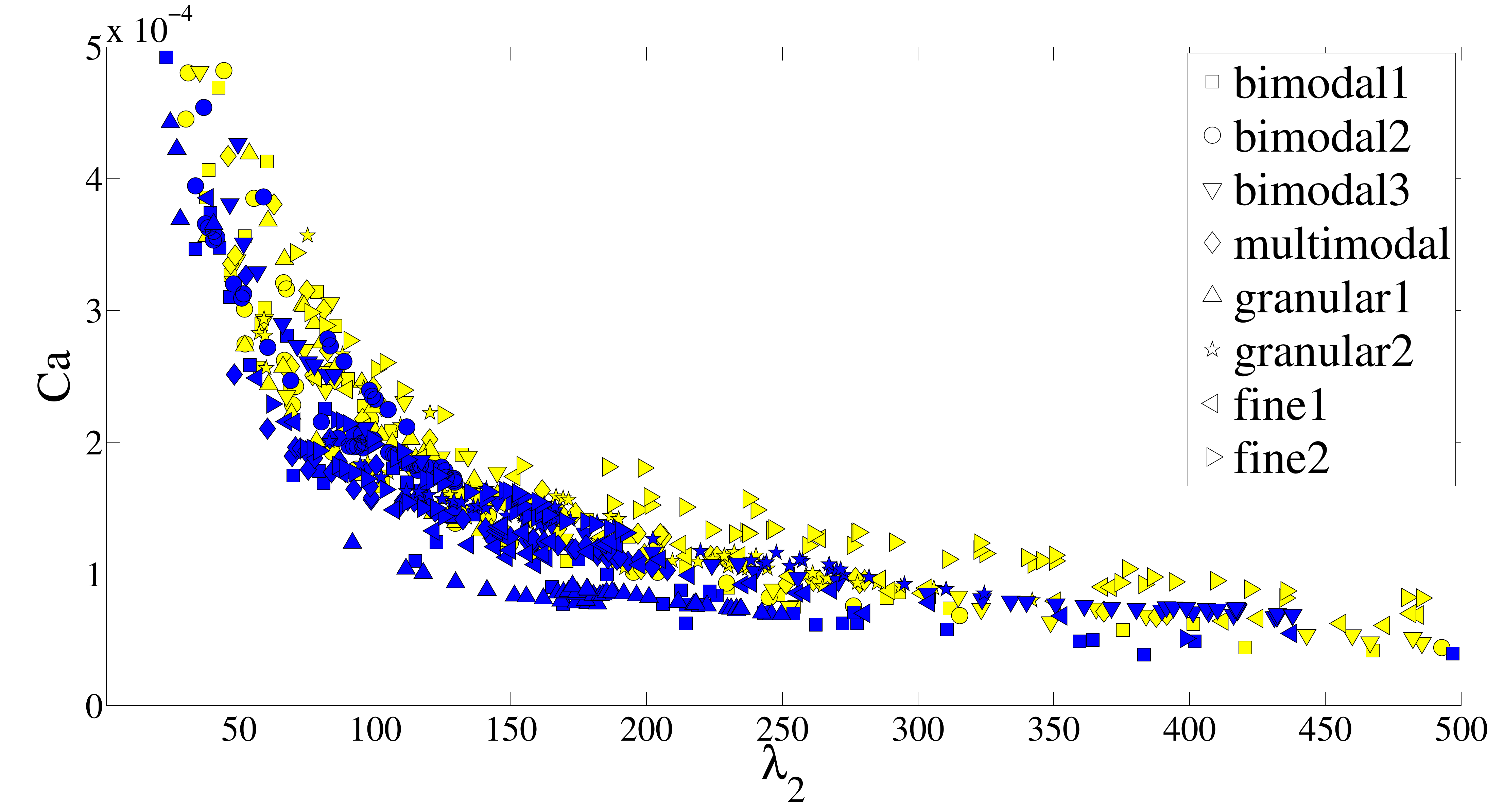}
\caption{\label{fig:lambda2}
Relationship between the capillary number and the tortuosity as obtained from the process of capillary
infiltration in the absence of reactivity. The packing structures are composed of spheres for yellow
markers and of aligned rhombs for blue markers. The properties of the packing systems are summarized
in Tabs.~\ref{tab:sphere} and \ref{tab:rhomb}.}
\end{figure}
\begin{figure}[t]
\includegraphics[width=8.4cm]{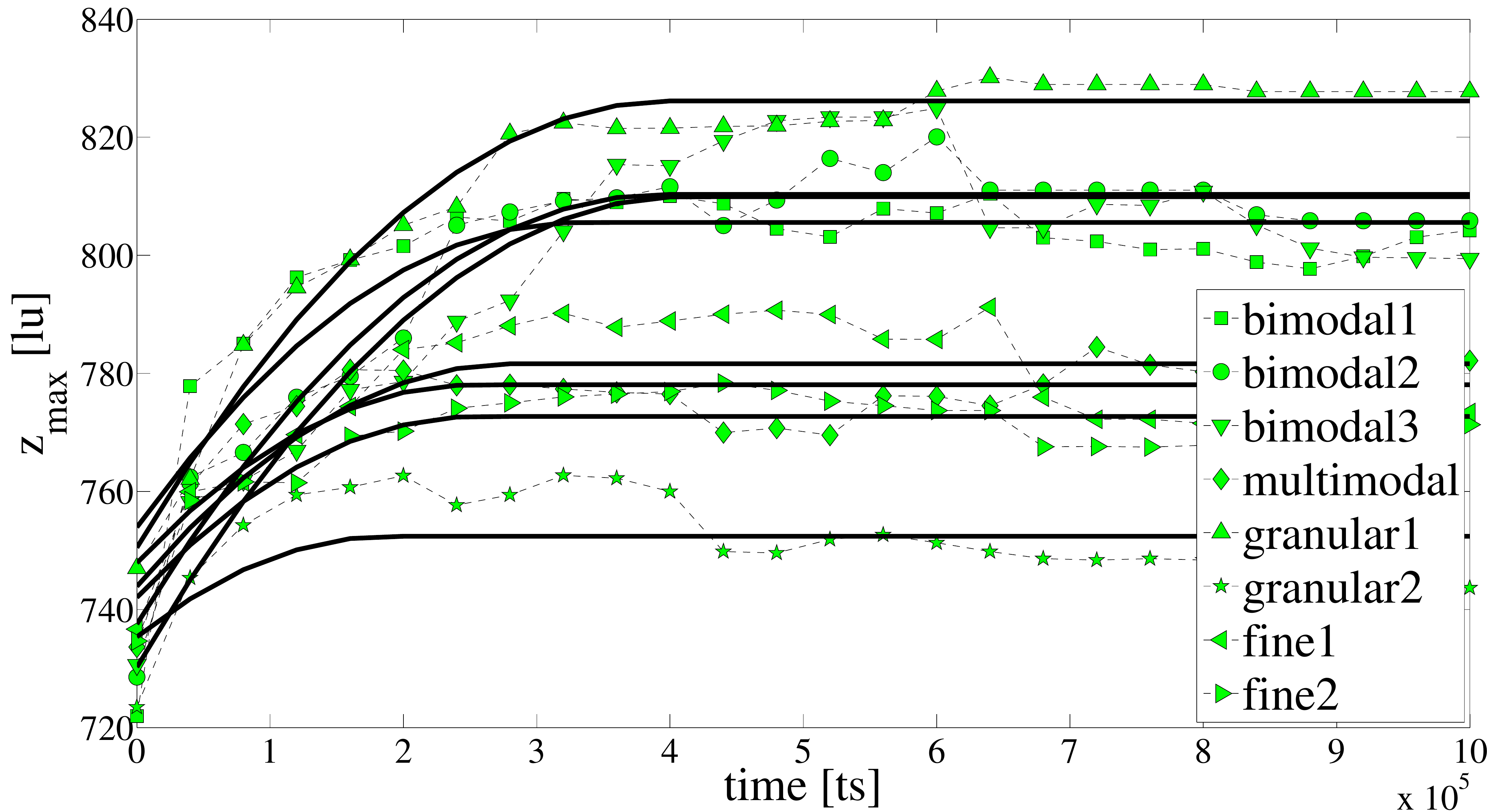}
\caption{\label{fig:zmax}
Average maximum of the invading front as time goes on for infiltrations with surface reaction. The porous
structures are obtained from the packing of rhombs with aligned particles (see Tab.~\ref{tab:rhomb}).
Points represent simulation results. The solid lines are fits to the data by means of Eq.~\ref{eq:reac}.}
\end{figure}
\begin{figure}[t]
\includegraphics[width=8.4cm]{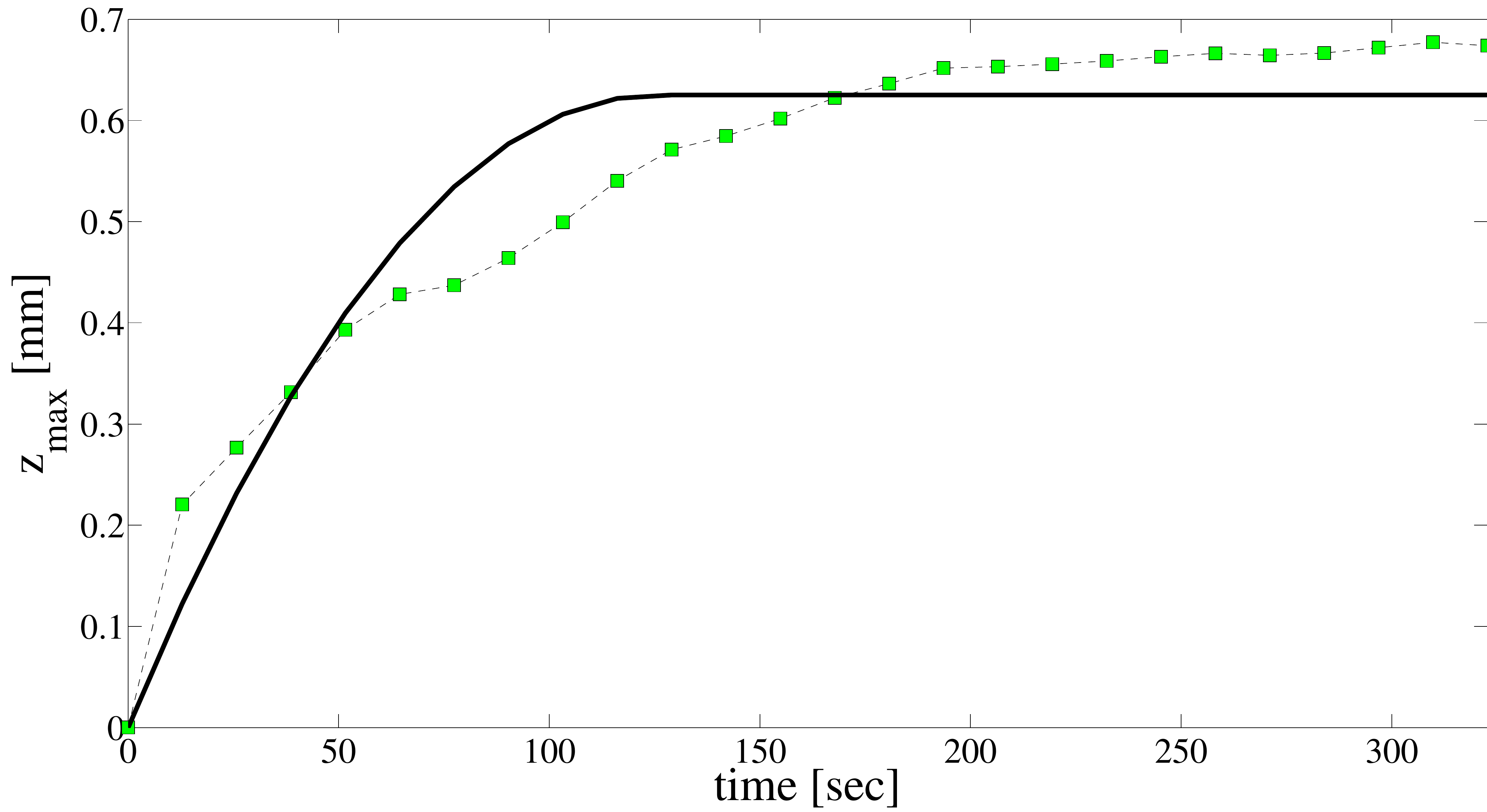}
\caption{\label{fig:zexp}
LB simulations for experimental results with surface growth. Average maximum for the infiltrated length
as a function of time. Points are used for results from simulations. The solid line is a fit to the data
with Eq.~\ref{eq:reac}.}
\end{figure}

\begin{sidewaystable*}[b]\vspace{7.5cm}
\begin{tabular}{lcccccccccc}
\hline\hline
Type  & $z\pm\Delta z$ [lu] & $v_{\mathrm{inf},1}$ $(10^{-4})$ [lu/ts] & $v_{\mathrm{inf},2}$ $(10^{-4})$ [lu/ts] & $r_{\mathrm{eff}}$ [lu] & $K$ [lu$^{2}$] & $\lambda_{1}$ & $\lambda_{2}$ $(10^{3})$ & $Ca$ $(10^{-4})$ & $Q_{4}$ $(10^{-8})$ & $Re$ $(10^{-4})$\\
\hline
bimodal1  & $915\pm 53$  & $0.71$ & $0.83$ & $0.44$ & $3.19$ & $1.17$ & $2.07$ & $1.40$ & $2.59$ & $1.85$\\
bimodal2  & $851\pm 60$  & $0.65$ & $0.51$ & $0.27$ & $4.65$ & $1.26$ & $0.56$ & $1.28$ & $1.35$ & $1.05$\\
bimodal3  & $842\pm 142$ & $0.65$ & $0.45$ & $0.24$ & $3.53$ & $1.22$ & $0.34$ & $1.29$ & $1.19$ & $0.92$\\
multimodal& $852\pm 70$  & $0.79$ & $0.52$ & $0.27$ & $2.22$ & $1.24$ & $0.26$ & $1.58$ & $2.06$ & $1.30$\\
granular1 & $814\pm 52$  & $1.09$ & $0.33$ & $0.17$ & $2.10$ & $1.34$ & $0.11$ & $2.16$ & $2.46$ & $1.14$\\
granular2 & $838\pm 107$ & $0.75$ & $0.41$ & $0.21$ & $1.80$ & $1.17$ & $0.19$ & $1.49$ & $1.44$ & $0.97$\\
fine1     & $790\pm 73$  & $0.41$ & $0.18$ & $0.09$ & $1.08$ & $1.30$ & $0.90$ & $0.82$ & $0.19$ & $0.23$\\
fine2     & $827\pm 100$ & $0.74$ & $0.34$ & $0.18$ & $0.97$ & $1.18$ & $0.27$ & $1.46$ & $1.17$ & $0.80$\\
\hline\hline
\end{tabular}
\caption{\label{tab:sphere_result}
Spheres composing the packing structures infiltrated without surface growth (see Tab.~\ref{tab:sphere}).}\vspace{0.5cm}
\hspace{-2.7cm}\footnotesize{}\begin{tabular}{lcccccccccccc} 
\hline\hline
Type      & \multicolumn{3}{c}{$z\pm\Delta z$ [lu]} & $v_{\mathrm{inf},1}$ $(10^{-4})$ [lu/ts] & $v_{\mathrm{inf},2}$ $(10^{-4})$ [lu/ts] & $r_{\mathrm{eff}}$ [lu] & $K$ [lu$^{2}$] & $\lambda_{1}$ & $\lambda_{2}$ $(10^{3})$ & $Ca$ $(10^{-4})$ & $Q_{4}$ $(10^{-8})$ & $Re$ $(10^{-4})$\\
\hline
bimodal1  & $878\pm 65 $ & $784\pm 99 $ & $876\pm  117$ & $0.73/0.72/1.04$ & $0.70/0.26/0.74$ & $0.37/0.14/0.39$ & $5.53/0.59/4.46$ & $1.15/1.18/1.20$ & $0.19/0.15/0.14$ & $1.45/1.41/2.06$ & $2.34/0.84/5.04$ & $1.62/0.60/2.45$\\
bimodal2  & $815\pm 76 $ & $901\pm 154$ & $849\pm  31$  & $1.29/1.27/0.64$ & $0.31/0.82/0.55$ & $0.16/0.43/0.29$ & $3.25/2.44/3.12$ & $1.31/1.29/1.34$ & $0.09/0.09/0.22$ & $2.56/2.52/1.27$ & $3.26/8.28/1.40$ & $1.27/3.28/1.11$\\
bimodal3  & $805\pm 74 $ & $835\pm 97 $ & $931\pm  151$ & $0.73/0.79/1.08$ & $0.26/0.44/0.95$ & $0.14/0.23/0.50$ & $1.01/2.13/4.59$ & $1.24/1.20/1.21$ & $0.27/0.15/0.20$ & $1.45/1.56/2.14$ & $0.87/1.70/6.96$ & $0.60/1.09/3.26$\\
multimodal& $789\pm 75 $ & $920\pm 72 $ & $840\pm  57$  & $0.76/0.99/0.44$ & $0.22/0.90/0.60$ & $0.12/0.47/0.31$ & $1.05/6.34/3.04$ & $1.22/1.18/1.32$ & $0.13/0.18/2.93$ & $1.51/1.96/0.88$ & $0.80/5.48/0.74$ & $0.53/2.80/0.84$\\
granular1 & $813\pm 93 $ & $815\pm 105$ & $811\pm  80$  & $0.70/0.64/0.57$ & $0.17/0.37/0.29$ & $0.09/0.19/0.15$ & $8.38/6.53/6.77$ & $1.26/1.10/1.28$ & $0.17/0.19/0.26$ & $1.38/1.28/1.13$ & $0.51/0.96/0.58$ & $0.37/0.75/0.52$\\
granular2 & $759\pm 91 $ & $835\pm 155$ & $1014\pm 213$ & $0.68/0.48/1.24$ & $0.13/0.39/1.19$ & $0.07/0.21/0.63$ & $0.57/3.19/12.3$ & $1.21/1.22/1.22$ & $0.19/0.35/0.20$ & $1.34/0.95/2.45$ & $0.37/0.56/1.14$ & $0.27/0.59/4.66$\\
fine1     & $799\pm 100$ & $801\pm 56 $ & $873\pm  62$  & $0.56/0.79/0.65$ & $0.27/0.26/0.57$ & $0.14/0.14/0.30$ & $1.42/2.18/10.1$ & $1.22/1.23/1.19$ & $0.32/0.17/0.43$ & $1.10/1.56/1.29$ & $0.53/1.03/1.51$ & $0.48/0.66/1.17$\\
fine2     & $799\pm 59 $ & $781\pm 99 $ & $818\pm  134$ & $0.84/0.29/0.43$ & $0.28/0.18/0.25$ & $0.15/0.10/0.13$ & $1.61/2.62/5.42$ & $1.21/1.21/1.20$ & $0.13/0.68/3.71$ & $1.65/0.58/0.85$ & $1.25/0.10/0.29$ & $0.76/0.17/0.35$\\
\hline\hline
\end{tabular}
\caption{\label{tab:rhomb_result}
Rhombs composing the packing systems infiltrated with inert boundaries (see Tab.~\ref{tab:rhomb}).
For every quantity, multiple values corresponding to different orientations of fillers: aligned, misaligned, misaligned
and tilted (see Fig.~\ref{fig:section}).}\vspace{0.5cm}
\hspace{-2.0cm}\begin{tabular}{lcccccccccccc}
\hline\hline
Type  & \multicolumn{3}{c}{$z\pm\Delta z$ [lu]} & $v_{\mathrm{inf},1}$ $(10^{-4})$ [lu/ts] & $v_{\mathrm{inf},2}$ $(10^{-4})$ [lu/ts] & $r_{\mathrm{eff}}$ [lu] & $K$ [lu$^{2}$] & $\lambda_{1}$ & $\lambda_{2}$ $(10^{3})$ & $Ca$ $(10^{-4})$ & $Q_{4}$ $(10^{-8})$ & $Re$ $(10^{-4})$\\
\hline
fiber1 & $920\pm 224$ & $828\pm 180$ & $1013\pm 124$ & $0.75/1.13/1.64$ & $0.81/0.42/1.38$ & $0.43/0.22/0.73$ & $8.24/3.27/9.00$ & $1.18/1.38/1.15$ & $0.52/0.08/0.07$ & $1.48/2.21/3.27$ & $2.83/3.32/23.5$ & $1.91/1.51/7.19$\\
fiber2 & $909\pm 79$  & $817\pm 101$ & $973\pm  138$ & $1.27/0.75/1.09$ & $0.78/0.39/1.01$ & $0.41/0.21/0.54$ & $4.29/1.28/6.72$ & $1.19/1.23/1.18$ & $0.10/0.16/0.16$ & $2.53/1.49/2.16$ & $7.96/1.40/7.55$ & $3.15/0.94/3.50$\\
fiber3 & $945\pm 326$ & $791\pm 125$ & $851\pm  64$  & $1.01/0.53/0.61$ & $0.92/0.24/0.60$ & $0.49/0.13/0.31$ & $5.25/1.49/5.02$ & $1.10/1.28/1.21$ & $0.15/0.25/7.13$ & $1.99/1.04/1.21$ & $5.84/0.42/1.39$ & $2.93/0.40/1.15$\\
fiber4 & $910\pm 278$ & $783\pm 59$  & $846\pm  48$  & $0.58/0.22/0.60$ & $0.86/0.12/0.32$ & $0.46/0.06/0.17$ & $2.30/4.98/12.3$ & $1.18/1.09/1.12$ & $1.10/0.50/0.97$ & $1.14/0.43/1.18$ & $1.79/0.04/0.72$ & $1.57/0.08/0.61$\\
fiber5 & $851\pm 204$ & $801\pm 91$  & $780\pm  98$  & $0.45/0.41/0.52$ & $0.48/0.32/0.18$ & $0.25/0.17/0.10$ & $5.41/3.73/4.11$ & $1.15/1.19/1.22$ & $0.38/0.57/0.27$ & $0.89/0.79/1.03$ & $0.61/0.33/0.31$ & $0.69/0.41/0.30$\\
fiber6 & $966\pm 205$ & $765\pm 77$  & $823\pm  57$  & $1.41/0.17/0.53$ & $1.12/0.09/0.29$ & $0.59/0.05/0.15$ & $7.67/2.67/3.28$ & $1.07/1.33/1.21$ & $0.15/4.29/2.12$ & $2.80/0.33/1.05$ & $14.1/0.02/0.50$ & $5.02/0.05/0.48$\\
fiber7 & $834\pm 75$  & $817\pm 78$  & $796\pm  88$  & $0.72/0.47/0.73$ & $0.37/0.37/0.23$ & $0.20/0.20/0.12$ & $2.66/1.28/5.19$ & $1.23/1.23/1.19$ & $0.18/0.33/0.19$ & $1.43/0.91/1.45$ & $1.21/0.50/0.77$ & $0.85/0.55/0.53$\\
fiber8 & $778\pm 81$  & $943\pm 188$ & $850\pm  53$  & $0.92/1.81/1.25$ & $0.17/0.87/0.61$ & $0.09/0.46/0.32$ & $2.49/5.22/2.10$ & $1.13/1.09/1.17$ & $0.12/0.07/0.08$ & $1.82/3.58/2.44$ & $0.92/17.8/5.86$ & $0.51/4.96/2.40$\\
fiber9 & $810\pm 133$ & $978\pm 160$ & $783\pm  150$ & $0.83/1.33/0.26$ & $0.33/1.04/0.15$ & $0.17/0.55/0.08$ & $2.59/5.82/2.31$ & $1.20/1.20/1.19$ & $0.15/0.09/3.10$ & $1.66/2.65/0.52$ & $1.44/11.6/0.06$ & $0.87/4.37/0.12$\\
\hline\hline
\end{tabular}
\caption{\label{tab:fiber_result}
Fibers added to the packing systems infiltrated without reactive boundaries (see Tab.~\ref{tab:fiber}).
For every quantity, multiple values refer to different orientations of fillers: aligned, misaligned, misaligned and tilted (see Fig.~\ref{fig:section}).}
\end{sidewaystable*}
\begin{sidewaystable*}[b]\vspace{7.5cm}
\begin{tabular}{lccccccccccc}
\hline\hline
Type      & $z_{\mathrm{max}}$ [lu] & $v_{\mathrm{inf},1}$ $(10^{-4})$ [lu/ts] & $r_{\mathrm{eff}}$ [lu] & $Q_{1}$ $(10^{-8})$ & $Q_{2}$ $(10^{-5})$ & $Q_{3}$ $(10^{-2})$ & $Ca$ $(10^{-4})$ & $Q_{4}$ $(10^{-8})$ & $Re$ $(10^{-4})$\\
\hline                    
bimodal1  & $810/781/793$ & $0.73/0.91/1.33$ & $0.83/0.54/0.77$ & $0.36/0.29/0.61$ & $2.48/1.61/2.32$ & $6.82/5.47/3.77$ & $1.45/1.78/2.61$ & $5.28/5.22/16.1$ & $3.63/2.94/6.16$\\
bimodal2  & $812/839/804$ & $1.57/1.08/0.54$ & $0.98/1.38/0.91$ & $0.91/0.88/0.29$ & $2.94/4.14/2.73$ & $3.19/4.65/9.24$ & $3.10/2.11/1.05$ & $28.7/18.8/3.10$ & $9.23/8.91/2.96$\\
bimodal3  & $815/810/846$ & $1.46/1.13/1.15$ & $1.03/1.00/1.49$ & $0.89/0.67/1.02$ & $3.08/3.01/4.47$ & $3.44/4.43/4.33$ & $2.89/2.23/2.28$ & $25.9/15.2/23.5$ & $8.96/6.79/10.3$\\
multimodal& $778/848/780$ & $0.99/1.88/0.70$ & $0.63/1.18/0.69$ & $0.37/1.32/0.29$ & $1.89/3.53/2.06$ & $5.05/2.65/7.18$ & $1.94/3.75/1.38$ & $7.27/49.9/3.97$ & $3.75/13.3/2.87$\\
granular1 & $822/827/826$ & $1.32/1.15/1.37$ & $1.00/0.82/1.12$ & $0.78/0.56/0.91$ & $3.00/2.45/3.35$ & $3.78/4.34/3.64$ & $2.61/2.30/2.72$ & $20.6/13.0/25.1$ & $7.92/5.65/9.21$\\
granular2 & $763/783/855$ & $0.88/0.61/1.04$ & $0.47/0.61/1.52$ & $0.25/0.22/0.94$ & $1.42/1.83/4.56$ & $5.67/8.21/4.82$ & $1.74/1.20/2.06$ & $4.36/2.69/19.5$ & $2.51/2.23/9.47$\\
fine1     & $788/789/863$ & $0.99/1.16/1.27$ & $0.71/0.82/1.25$ & $0.41/0.57/0.93$ & $2.12/2.47/3.74$ & $5.07/4.31/3.94$ & $1.93/2.29/2.49$ & $8.06/13.1/23.6$ & $4.18/5.73/9.48$\\
fine2     & $774/781/833$ & $0.89/0.66/1.05$ & $0.64/0.64/1.19$ & $0.32/0.25/0.74$ & $1.91/1.92/3.57$ & $5.63/7.61/4.78$ & $1.70/1.30/2.07$ & $5.76/3.27/15.5$ & $3.39/2.52/7.47$\\
\hline\hline
\end{tabular}
\caption{\label{tab:reaction_result}
Surface reaction enabled for the infiltration of packing systems obtained from rhombs (see Tabs.~\ref{tab:rhomb} and \ref{tab:rhomb_result}). For every quantity, multiple
values are associated with different orientations of the fillers: aligned, misaligned, misaligned and tilted (see Fig.~\ref{fig:section}).}
\vspace{0.5cm}
\end{sidewaystable*}
\begin{table*}[t]
\begin{ruledtabular}
\begin{tabular}{lcccccc}
                 & $Q_{1}$             & $Q_{2}$             & $Q_{3}$            & $Ca$               & $Q_{4}$             & $Re$ \\
\hline
LB systems       & $2.55\cdot10^{-9}$  & $1.45\cdot 10^{-5}$ & $2.81\cdot 10^{-2}$ & $1.76\cdot 10^{-4}$ & $9.06\cdot 10^{-8}$ & $5.16\cdot 10^{-4}$ \\
real systems     & $1.48\cdot10^{-16}$ & $2.71\cdot 10^{-8}$ & $0.80\cdot 10^{-2}$ & $5.47\cdot 10^{-9}$ & $1.85\cdot 10^{-14}$ & $3.39\cdot 10^{-6}$
\end{tabular}
\end{ruledtabular}
\caption{\label{tab:exp_result}
Characteristic dimensionless numbers for infiltrations in the presence of surface growth. The results for LB simulations are obtained from the
systems with higher resolution leading to Fig.~\ref{fig:zexp}. In the case of real systems, typical experimental parameters are used
(see Sec.~5.2).}
\end{table*}

\section*{3.~~~PACKING SYSTEMS}

It is standard practice to reproduce the microstructure of composite materials with random packings
(Scocchi et al., 2013; Torquato, 2002). The problem consists in filling the empty space with particles arranged in
a disordered way. The realization of high volume fractions can also be quite challenging (Donev et al., 2004).
The principle of the algorithm for the random sequential addition is to place one particle after the other
in the simulation domain (Sherwood, 1997; Widom, 1966). In case of overlapping between a test particle and an
existing one, the trial is not accepted and a new particle is considered. This procedure is iterated up to the
desired filling fraction. With this algorithm it is not easy to exceed values beyond $0.6$.

In order to generate the packing systems for the LB simulations we employ the algorithm proposed in the
article by Sergi, D'Angelo, Scocchi, and Ortona (2012). The method consists in decomposing the particles into small cubes.
The lengths are measured in units of the small cubes composing the particles according to the rule
$\ell=(n+1)r$. $n$ is the number of cubes and $2r$ their side length. Here, we consider spheres, rhombs and sticks
with a certain degree of randomness for their size. For a given value of $n$, the associated lengths are
chosen with a Gauss distribution of average $\ell$ and standard deviation $\sigma=\ell/4$. When the rotation
of the particles is allowed, the rotation angles around the three axes are at most
$\theta_{x}=\theta_{y}=\pi/2$ and $\theta_{z}=\pi/4$.
The domain is a cube of side $1$ divided into $N=256$ bins. The porous medium for the LB simulations is
obtained by suitably replicating the packing systems. The minimal distance between two small cubes is
$d_{\mathrm{min}}=2/N$. An external pressure is applied toward the center of the domain in order to obtain
more compact structures when necessary, as it is the case for higher degrees of disorder, for example.
The other parameters for the pressure are the same as in Sergi et al.~(2012).

The packing systems aim at reproducing basic characteristics resulting from the use of common commercial
powders for preform preparation. In order to make the distinction between round-shaped and sharp-edged
morphologies for the particles, we consider spheres and rhombs. For example, the latter geometry can approximate
the elongated shape of SiC particles. Fibers are also used as an attempt to understand the role of long walls.
Concerning the size, the distribution is bimodal, multimodal or monomodal (coarse and fine particles).
The properties of the packing systems are summarized in Tabs.~\ref{tab:sphere}-\ref{tab:fiber}. Figure
\ref{fig:section} shows the section for some systems.


\section*{4.~~~SETTINGS FOR LB SIMULATIONS}\label{sec:sim}

As said before, the LB method is based on the discretization of the velocity space and a statistical treatment of the particle
motion. In the BGK approximation for the collisions (Bhatnagar, Gross, \& Krook, 1954), LB simulations remain versatile, efficient and accurate.
Proper hydrodynamic behavior can be recovered in the incompressible limit at low Mach numbers (Chen \& Doolen, 1998).
The domain for simulations is $N_{x}=1500$ lu long and $N_{y}=255$ lu wide. The porous structure has a length of $L=N_{x}/2$ and
the solid-liquid interface starts at $x=2N_{x}/5$. The total number of timesteps is given by $N_{t}=2\cdot10^{6}$ ts. In the region
occupied by the porous structure the boundaries of the simulation domain are solid and subject to the common bounce-back rule;
elsewhere they are periodic. This is done in order to reduce the phenomenon of pinning near narrow-to-wide parts
(Blow, Kusumaatmaja, \& Yeomans, 2009; Chibbaro, Biferale, Binder, et al., 2009;
Chibbaro, Costa, et al., 2009;
Kusumaatmaja, Pooley, Girardo, Pisignano, \& Yeomans, 2008; Mognetti \& Yeomans, 2009;
Wiklund \& Uesaka, 2012, 2013).
The reason is that pinning can result in a significant slowdown of infiltration or even stop the flow.
As explained before, the problem is studied with a two-component system, or binary mixture
(Chibbaro, 2008; Chibbaro, Biferale, Diotallevi, et al., 2009;
Sukop \& Thorne, 2010). The viscosity is kept fixed with a relaxation
time of $\tau=1$ ts for both components. This choice is in general a guarantee for numerical stability (Sukop \& Thorne, 2010).
In the initial condition, half of the simulation domain is filled with
the wetting fluid and the remaining with the non-wetting one. At start, the density for the main component is $\rho=1.95$
mu/lu$^{2}$ and that for the dissolved component $\rho=0.05$ mu/lu$^{2}$. The parameter for fluid-fluid interactions
(cohesive forces) is set to $G_{\mathrm{c}}=0.9$ lu/mu/ts$^{2}$ (Martys \& Chen, 1996;
Sukop \& Thorne, 2010), resulting in a surface tension of
$\gamma=0.16403$ lu$\cdot$mu/ts$^{2}$ (Sergi et al., 2014). Solid-fluid interactions (adhesive forces) are determined by
the parameters $G_{\mathrm{ads},1}=-G_{\mathrm{ads},2}=-0.35$ lu/ts$^{2}$ (Martys \& Chen, 1996;
Sukop \& Thorne, 2010). With this choice the equilibrium
contact angle turns out to be $\theta=30^{\circ}$ (Sergi et al., 2014). This value is typical for Si droplets on SiC
(Bougiouri et al., 2006; Voytovych et al., 2008). The details of the LB models can be found in Sergi et al. (2014).

The process of surface growth from the reaction is introduced by relying on a purely heuristic treatment. A description based
on solute transport as in Sergi et al.~(2014, 2015) exhibits limitations related to the interface for solute transport. Namely,
inside the porous structures the interface can form, leading to the diffusion of solute in the whole system. Therefore, in this
work the reaction is implemented by making grow the surface of the solid phase behind the contact line at regular intervals
of time. Precisely, the first layer around the solid boundaries is converted to solid phase after every $N_{t}/10$
timesteps. The reaction-rate constant is thus given by $k=5\cdot10^{-6}$ lu/ts. This framework has the advantage to be suitable
for a critical volume of simulations allowing to observe effects of pore structure and surface growth on fluid flow. Unless
stated otherwise, these settings are maintained in the sequel.


\section*{5.~~~RESULTS AND DISCUSSION}

\subsection*{5.1~~~Packing systems}

In Tab.~\ref{tab:sphere_result} we report the results of simulations for systems obtained from the packing of spheres
in the absence of surface reaction. Here $z$ indicates the penetration attained on average at the end of the process.
$\Delta z$ is an estimate of the deviations from the average. This quantity is determined from the maximum and the minimum
of the front displacement. It is to be noted that the error $\Delta z$ is comparable to the infiltrated distance. This means
that the fluid advances into the porous medium through selected pathways (see Fig.~\ref{fig:front}). It can be expected that these fingers
then contribute to fill the pores left behind. It is worth noticing that the average error is $82$ lu while the size of
the largest particles is on average of $51$ lu. In Fig.~\ref{fig:filling1} is shown the kinetics of invasion.
It turns out that the infiltration is slower for a monomodal distribution of the particle size. By fitting the data with
Eq.~\ref{eq:linear}, we derive the infiltration velocity $v_{\mathrm{inf},2}$ and the effective radius $r_{\mathrm{eff}}$ (see
Tab.~\ref{tab:sphere_result}). It can be seen that the effective radius is one to two orders of magnitude smaller than
the average particle size, as stressed in previous works (Einset, 1996; Patro, Bhattacharyya, \& Jayram, 2007). 
The permeability $K$ is determined under the conditions usually employed to reproduce Poiseuille flow,
i.e.~with a single fluid, periodic boundaries and an external acceleration (Sukop \& Thorne, 2010).
In Tab.~\ref{tab:sphere_result} it can be seen that the permeability for spheres is higher with bimodal size distributions.
In order to assess the role
of interface dynamics, the tortuosity is computed in two different ways. $\lambda_{1}$ is determined
in the presence of an external acceleration as for the permeability. 
$\lambda_{2}$ is instead obtained from the process of infiltration driven by capillary forces.
The results of Tab.~\ref{tab:sphere_result} show that $\lambda_{2}$ is two orders of magnitude higher. As a consequence,
the interface dynamics induced by the pore structure plays a prominent role. $v_{\mathrm{inf},1}$ is obtained by an average
over the velocities inside the porous media in the direction of infiltration, i.e.~$x$. The difference with $v_{\mathrm{inf},2}$
is more marked when the process is slower. For a proper basis of comparison with systems including the reaction, the
characteristic dimensionless numbers are calculated by means of $v_{\mathrm{inf},1}$ (see Tab.~\ref{tab:sphere_result}).
It is also interesting to consider the Mach number defined as $Ma=v/c_{\mathrm{s}}$, where $c_{\mathrm{s}}=\sqrt{1/3}$
lu/ts is the speed of sound. From Tab.~\ref{tab:sphere_result} it can be seen that the order of magnitude of  $Ma$ is $10^{-4}$.
It follows that in this study the numerical error associated with the LB scheme remains in the tolerance limits (He \& Luo, 1997;
Sukop \& Thorne, 2010).

The results of simulations without reactivity for packings of particles with the shape of rhombs are reported in
Tab.~\ref{tab:rhomb_result}. Figure \ref{fig:filling1} shows the kinetics in the case of aligned rhombs. Comparison with
the outcome for packings of spheres indicates that the structure arising from rhombs is more subject to pinning. It turns
out that, as the degree of disorder increases, the infiltration becomes faster. The fastest infiltrations are obtained with
tilt and misalignment, even though the tortuosity $\lambda_{2}$ attains a larger value on average. The dynamics with spheres
remain still faster or at least comparable. A possible explanation for this observation is that the arrangement of rhombs can
give rise to channels that are filled easily. The shortcoming is that the phenomenon of pinning becomes more frequent. In 3D
in a real preform, we expect that the advantages should prevail since
the connectivity of the pore chambers is higher and the number of pathways increases. For example, the
particle walls above or below a 2D section and almost parallel to it contribute to reduce pinning. This point will become
clearer from the discussion on the advantages of bimodal size distributions.
As expected, the permeability $K$ is higher with increasing orientation disorder
(see Tab.~\ref{tab:rhomb_result}). For aligned systems it also turns out that $K$ attains higher values in the presence of
larger particles.
Figure \ref{fig:q4} shows the range of variation of $Ca$. The same behavior
holds for $Re$: the same relation between $Re$ and $Ca$ is of course an identity. More marked variations are observed for $Q_{4}$.
The increase can amount to one order of magnitude. It is interesting to note that the deviations from the average values
are stronger for rhombs since the filling of distinct channels and depinning introduce more accelerations. From the order of magnitude
of the average values of $Ca$, $Q_{4}$ and $Re$, we conclude that the effects of inertia are weak and capillary forces remain
the dominant ones. Furthermore, in Fig.~\ref{fig:lambda2} it can be seen that for a given capillary number, the tortuosity seems to be
slightly lower for aligned rhombs in a comparison with spheres. This tendency is recognized also in the presence of orientation
disorder and for fluctuations of $Ca$. For clarity, this suggests that, when the fluid accelerates, the tortuosity is not higher and
thus the resistance of inertia has a weak effect. So, in this case the averages of dimensionless numbers still provide
an exhaustive description. As expected, it follows that the interface dynamics induced by the pore structure seems to
play a more important role for the pronounced difference between $\lambda_{1}$ and $\lambda_{2}$.
Spurious currents at the interface should average out since we could verify that, when the fits allowing
to obtain $v_{\mathrm{inf},2}$ are more precise, the discrepancies with $v_{\mathrm{inf,1}}$ are smaller. Again, for aligned rhombs,
it arises that liquid penetration tends to be better with a bimodal distribution. This trend is less clear with misalignment
and tilt. It should be said that, in this case, 2D sections loose to a large extent the specific characteristics of the particles
composing the packings. For this reason, it might be presumed that, in 3D, bimodal distributions should remain more advantageous.
Examination of the data associated with the maxima and minima for infiltration reveals that pinning is the discriminating factor
coming into play. The basic mechanism at the origin of pinning is that the meniscus advances slowly while loosing its curvature,
resulting in a drop of capillary pressure. In the presence of channel walls, the front can travel a longer distance since the
pinning sites are not vertically aligned. As a consequence, the meniscus can encounter the surface of new particles with higher
probability before halting (see Fig.~\ref{fig:front}). In the packings with small particles the pinning sites are instead generally separated by small
distances. The probability that the invading front stops moving is thus higher. It is also interesting to note that rhombs
with bimodal distributions pack more efficiently. On average the filling fraction of their 2D sections is $14\%$ higher than
in the case of monomodal distributions. However, their infiltrations remain faster on average of $3\%$. It can be seen that
the results are more sensitive to specific porous configurations than to the porosity. For more clarity, test simulations are
performed for different porosities for the three bimodal systems  composed of rhombs with particles aligned and misaligned.
For filling fractions of $36\%-50\%$ the penetration depth varies within a narrow range on average. This is partly in line with our discussion
for the conditions helping the advancement of fluid. The porosity seems to start becoming detrimental for a filling fraction
approaching $60\%$.

Packing structures with sticks are also considered in order to make clearer the role of channels arising from the
arrangement of particles introducing long walls. Table \ref{tab:fiber_result} summarizes the results of simulations.
It is manifest that the systems with alignment lead to the fastest infiltrations. Furthermore, within this group
the dynamics tend to be faster in the presence of longer sticks. These results show that fibers favor the formation
of preferential infiltration paths. This comes out also from the high values of the error $\Delta z$ on the average
penetration depth. With orientation disorder, results of rhombs and fibers are still comparable (cf.~Tab.~\ref{tab:rhomb_result}).
Sticks introduce more channels that are filled easily, inducing
stronger accelerations. This appears also from the fluctuations of $Ca$ and $Re$ that turn out to be more marked, displaying
also higher peaks (results not shown for brevity). Nevertheless, also for this set of data, capillary forces remain the dominant
ones. It is interesting to note that the simulations with misalignment and tilt can lead to quite fast infiltrations, even though
the number of channels is lower. A possible explanation is that orientation disorder for fibers can be regarded as equivalent
to the presence of small particles when 2D cross-sections are considered (see Fig.~\ref{fig:front}). As a consequence, when the meniscus slides along
the walls introduced by fibers, it is likely to encounter the surface of other particles before the interface becomes flat
and pinning occurs. So, the results for fibers confirm the advantages identified for pore configurations associated with
bimodal size distributions. The permeabilities for packings with fibers are more informative on the performace
of the porous systems for capillary infiltration than in the presence of rhombs.
It is also interesting to consider the influence of the surface area of the solid phase (Bear, 1972), i.e. the length
of its boundary for 2D structures. In principle, for a given volume, spherical shapes tend to minimize the surface area
while thin oblate spheroids mazimize it. It is found that for systems obtained from spheres and rhombs, the boundary
length of the solid phase is higher for the systems composed of small particles, in particular for spheres.
The boundary of the solid phase is longer in the case of aligned fibers. With orientation disorder, the length of the
solid boundary is still longer than that of systems with rhombs. By considering for the solid phase its surface area divided
by its volume, known as specific surface (Bear, 1972), it is found that the systems with fibers exhibits higher values than
bimodal systems on average. The specific surface is in use for the characterization of powders. For our systems, the specific
surface is lower for rhombs because it is easier to realize compact packings with coarse particles. The highest values are
obtained with fibers that have however the advantage to favor the formation of channels.

Simulations in the presence of surface growth are performed for packing structures composed of rhombs (see Tabs.~\ref{tab:rhomb}
and \ref{tab:rhomb_result}). A typical configuration is shown in Fig.~\ref{fig:front}. Figure \ref{fig:zmax} shows examples of
infiltration kinetics. At the interface between the solid and the wetting fluid the front retreats as the surface grows. This effect
is more important when agglomerates of particles merge. In the analysis of the data, we thus consider an average over the highest
values for the position of the front. Fits to Eq.~\ref{eq:reac} allow to obtain the initial effective radius $r_{0}$. The effective
radius used in order to determine the characteristic numbers is set to the average value: $r_{\mathrm{eff}}=r_{0}/2$. The question arises whether
infiltrations stop because of pore closure or pinning. Examination of the fastest dynamics indicates that the cause due to
pore closure is as frequent as that for the barrier introduced by pinning. This implies that surface growth enhances the phenomenon
of pinning. Indeed, it is found that in the period considered for the fits, the infiltrations without surface reaction
are on average $5.7\%$ faster. The resistance of surface growth is more marked for the fastest dynamics, i.e.~with higher
orientation disorder. At the point where infiltration stops according to the theoretical model of Eq.~\ref{eq:reac}, the
results with surface growth are comparable to that with inert boundaries. But it is to be noted that $r_{0}$ turns out to
be underestimated in particular for the longer infiltrations where the front can overcome pinning barriers. More details for
the results are summarized in Tab.~\ref{tab:reaction_result}. It can be seen that orientation disorder is associated with
faster infiltrations. Bimodal size distributions lead to better results for aligned and misaligned particles. This is no
more true with the addition of tilt. In any case, it can be expected that, in 3D porous structures, bimodal size distributions
still present more advantages. In Tab.~\ref{tab:reaction_result}, the infiltration velocities are higher than in
Tab.~\ref{tab:rhomb_result}. With surface reaction, the dynamics is shorter and the initial fast infiltration has more weight.
It should be kept in mind that now the dominant effect is the infiltration interruption directly caused or induced by
surface growth. With surface growth allowed, $Ca$, $Q_{4}$ and $Re$ increase appreciably (see Tab.~\ref{tab:rhomb_result}).
Furthermore, their fluctuations are by a factor of $2$ smaller than for inert boundaries. $Q_{4}$ gains on average one order
of magnitude, but the effects of capillary and viscous forces remain more significant. This arises also from the characteristic
numbers involving the properties of surface reaction $Q_{1}$, $Q_{2}$ and $Q_{3}$. These quantities describe better the systems
because infiltration stops under the action of surface growth. $Q_{1}$ is more accurate because the effects of capillary forces
are second in importance even though their relevance weakens with respect to viscosity and inertia in the presence of surface
growth. Only the fluctuations of $Q_{3}$ in a few cases can deviate from the average value more than one order of magnitude.

\subsection*{5.2~~~Comparison with experiments}

We now want to carry out simulations for typical experimental results with pure Si (Eustathopoulos, 2015; Israel et al., 2010).
The initial porosity of the preform can be assumed to be $0.3$ with initial average pore diameter of $d=10$ $\mu$m. The pore size
distribution is assumed to be quite narrow, with pores mainly in the range $7-13$ $\mu$m (diameters). The preform is composed of
graphite particles with average size of $70$ $\mu$m. The reaction-rate constant is $k=4\cdot10^{-8}$ m/s (Messner \& Chiang, 1990).
The other parameters of Si are $\rho=2.53\cdot10^{3}$ kg/m$^{3}$ for the density, $\mu=0.94\cdot10^{-3}$ N$\cdot$s/m$^{2}$ for the
viscosity and $\gamma=0.86$ N/m for the surface tension (Einset, 1996).
For the infiltration velocity we use $v=5$ $\mu$m/s (Eustathopoulos, 2015). Thus, the holding time for the experiment can
be estimated to be $t_{\mathrm{pc}}=125$ sec. From the average pore radius it follows that at this time, indicated by $t_{\mathrm{pc}}$, pore
closure occurs and the infiltration stops. The maximal infiltration depth results to be $z_{\mathrm{max}}=0.625$ mm.
The effective radius is estimated using Dullien model as in the article by Einset (1996). By taking into account the thickening
of the surface of $2.5$ $\mu$m leading to the average pore diameter $d/2$, it is found that an estimate for the effective radius
is $r_{\mathrm{eff}}=0.25$ $\mu$m.
Indicatively, in industrial practice targeted by us (e.g., manufacturing of burners and radiation plates),
the C preforms are obtained starting from powders with bimodal size distributions. The size of the largest particles can vary
in the range of $30-70$ $\mu$m. The size of the smallest particles can be assumed to be $5$ $\mu$m. For example, typical
experimental work would proceed as described in the articles by Israel et al.~(2010) and Voytovych et al.~(2008).

Given the above experimental conditions, for the simulations we consider systems $N_{x}=1500$ lu long and
$N_{y}=511$ lu wide. Since the width
is twice longer than before, roughly speaking the infiltration velocity may double, as suggested by Eqs.~\ref{eq:reac} and
\ref{eq:linear}. The length of the samples is again $L=N_{x}/2$, placed in the simulation domain as described in Sec.~4. The total
number of timesteps is now set to $N_{\mathrm{t}}=4\cdot10^{6}$ ts. The surface grows after every $N_{\mathrm{t}}/10$ timesteps. In
simulations, the reaction-rate constant is thus given by $k=2.5\cdot10^{-6}$ lu/ts. All the other parameters for the LB simulations
remain unchanged. The porous structures are obtained from the packing of rhombs with average side and height corresponding to $36$ lu
and $31$ lu. Both misalignment and tilt are allowed.
The resolution of the systems does not allow to include small particles as in industrial applications.
The porosity of the resulting system is $0.64$. In this case, the sides of the
unit cube defining the porous structure are subdivided into $512$ bins. Simulations are performed for $30$ sections in 2D. The
correspondence with experimental results is established by imposing the experimental values for $z_{\mathrm{max}}$ and $t_{\mathrm{pc}}$.

Figure \ref{fig:zexp} shows the kinetics of infiltration as obtained from simulations. As done before, an average is taken over
the highest values for the positions of the fronts. It can be seen that the model predicts a more abrupt behavior for the interruption
of penetration, as observed also in experiments (Israel et al., 2010). It turns out that after an initial fast dynamics, the infiltration
velocity decreases and in the last stage the maximal depth is attained slowly. For the sake of clarity, we consider suitable averages
in the neighborhood of the point where pore closure occurs and in the ensuing interval of time where a plateau is expected to form. It
is found that for $63\%$ of the systems the infiltrated distance increases more than $10\%$ after the time $t_{\mathrm{pc}}$ where pore
closure is predicted by the model for the data of Fig.~\ref{fig:zexp}. On average, the further advancement amounts to $45\%$. For the
remaining $37\%$ of the systems after the time $t_{\mathrm{pc}}$ the fronts travel on average an additional distance corresponding to
$0.6\%$. It is verified that in this case the flow interruption can be ascribed to pinning. As a result, the value of $t_{\mathrm{pc}}$
determined by the model is practically associated with the first pinning barrier. When the plateau start forming the effect of surface
growth seems to remain mainly indirect. It is to be noted that the simulations reveal the tendency that the interruption of invasion
is affected by the combination of the effects of surface growth, tortuosity and interface dynamics. In principle, it is reasonable to
presume that the infiltration may stop before the time required for the obstruction of the average radius since the process of
infiltration is described by an effective radius. However, we find that in the simulations the effect of pinning is more critical
than expected in experiments. In order to elucidate the role of the effective radius more experiments are also needed for given
well-known reaction-rate constants (Israel et al., 2010).

In Tab.~\ref{tab:exp_result} the characteristic numbers of simulations and experiments are compared.
The objective is to assess the quality of the equivalence between experimental and simulation conditions.
It appears that the effects of capillary forces are too weak with respect to the effects of surface growth and viscous forces.
For the capillary number it is well known that high
values satisfy $Ca>10^{-2}$. For this reason, it could be questioned if the difference for the orders of magnitude of $Q_{1}$ is of
particular relevance. For sure, our approach allows to establish an equivalence for the dynamics of the average penetration depth.
Instead, the details of the invasion process can not be taken for granted. For example, for estimates based on the average
radius of the porous structure, in experiments, when the infiltration stops because of pore closure, the ratio of the infiltrated
distance to the thickness of the reaction-formed SiC is $125$. In simulations, it is obtained $75$.
As discussed above, the relation between the average and effective radii deserve attention but a major hurdle in order to achieve
better results is represented by the phenomenon of pinning. These simulations also prove that a higher resolution is not sufficient
in order to reduce pinning. Again, we see that the structure is more important than the average radius, of course increasing with
a higher resolution. Under the conditions of these simulations, the average size of the particles is $150$ $\mu$m while the target
was $70$ $\mu$m for an equivalence with a stronger direct correspondence for the porous structure. It is not easy to improve the
comparison because with packings composed of smaller particles, pinning would become even more important. Furthermore, if we use
the data associated with the first point in Fig.~\ref{fig:zexp} exceeding the average of $z_{\mathrm{max}}$ over the last $10$
frames taken into account for the fit, the improvement is relative. The reason is that the velocity decreases but the initial
effective radius $r_{0}$ increases. The result is that $Q_{1}$ and $Ca$ remain of the same order. Thus, it is not straightforward
in the simulations to reduce the velocity, to keep $r_{0}$ small and to weaken the phenomenon of pinning at the same time.


\section*{6.~~~CONCLUSIONS}

In this work, LB simulations are proposed for the study of capillary infiltration into porous structures. Optimization principles
for the manufacturing of ceramic materials through reactive melt infiltration are at the center of our attention. In this aim,
first are considered porous systems realized with the packing of particles with different properties (shape, size distribution,
orientation disorder). Without surface growth, our analysis shows that the invading front tend to select preferential pathways.
This can result in the formation of fingers. The data also indicate that bimodal size distributions display better infiltrations
than monomodal distributions. The improvement is found to be $4.1\%$ on average (see Tabs.~\ref{tab:sphere_result} and \ref{tab:rhomb_result}).
Pinning of the contact line appears as the primary factor affecting liquid penetration. Structural disorder can reduce its effect.
Interestingly, bimodal size distributions present advantages in order to limit pinning. Larger faceted particles introduce walls
giving rise to channels. As a result, the meniscus travels longer distances before the loss of its curvature. The probability to
encounter the surface of other particles is even higher with the addition of smaller particles. The results for packings with fibers
support this view. It is also verified that inertia effects induced by transient accelerations remain of secondary importance.
Surface growth results in a clear interruption of the flow. This can occur because pinning turns out to be enhanced or for
the pore-closing phenomenon. The retardation effects are more marked for the fast dynamics with a higher degree of orientation
disorder. Again, the results point out that bimodal size distributions appear to be able of accommodating optimal infiltration.
Simulations with higher resolution are also performed for a comparison with experiments. The relative strength of capillary
forces turns out not to be strong enough. As a result, the infiltration dynamics is not reproduced in full details.
Furthermore, the effects of surface growth for the interruption of infiltration are basically indirect in this case. We identify two
aspects significant for further advances. First, the relation for the interruption of infiltration with the physical (e.g., average
and minimum) and effective radii needs to be elucidated. Second, longer infiltrated lengths should be realized in simulations.
A major obstacle to improvements resides in the over-occurrence of pinning.
This situation could be improved by considering other combinations of surface tension, viscosity
and, especially, contact angle (Chibbaro, Costa, et al., 2009). This could also be an extension of this work to Si alloys.
Finally, our study develops a simulation work relevant for increasing the
quality of ceramics obtained from LSI. Despite a poor statistics and noisy results, operational guidelines are highlighted. Further
research is necessary in order to sharpen the conditions determining the optimal configuration for the porosity. Experimental tests
would also allow to understand the robustness of our findings based on simplified systems.


\acknowledgments

The research leading to these results has received funding from the European Union Seventh Framework Programme
(FP7/2007-2013) under grant agreement n$^{\circ}$ 280464, project "High-frequency ELectro-Magnetic technologies
for advanced processing of ceramic matrix composites and graphite expansion'' (HELM).


\end{document}